\documentclass[a4paper,11pt]{article}
\pdfoutput=1 

\usepackage{jinstpub} 

\usepackage[utf8]{inputenc}
\usepackage[english]{babel}

\usepackage{graphicx}
\usepackage{caption}
\usepackage{subcaption}
\usepackage{hyperref}
\usepackage{graphicx}
\usepackage{csquotes}
\usepackage{hyperref}

\usepackage{todonotes}

\title{Prototype of a segmented scintillator detector for particle flux measurements on spacecraft}

\author[1,2]{Egor Stadnichuk}
\author[1]{Tatyana Abramova}
\author[1,2,3]{Mikhail Zelenyi}
\author[2]{Alexander Izvestnyy}
\author[1,2]{Alexander Nozik}
\author[1,2]{Vladimir Palmin}
\author[3]{Ivan Zimovets}

\affiliation[1]{Moscow Institute of Physics and Technology (National Research University) - MIPT, 1 "A" Kerchenskaya st., Moscow, Russia, 117303}

\affiliation[2]{Institute for Nuclear Research of the Russian Academy of Sciences (INR RAS), Prospekt 60-letiya Oktyabrya 7a, Moscow, Russia, 117312}

\affiliation[3]{Space Research Institute of the Russian Academy of Sciences (IKI RAS), 84/32 Profsoyuznaya st., Moscow, Russia, 117997}

\emailAdd{yegor.stadnichuk@phystech.edu}

\abstract{
    In this paper, we introduce a laboratory prototype of a solar energetic particle (SEP) detector which will operate along with other space-based instruments to give us more insight into the SEP physics. The instrument is designed to detect protons and electrons with kinetic energies from 10 to 100 MeV and from 1 to 10 MeV respectively. The detector is based on a scintillation cylinder divided into separated disks to get more information about detected particles. Scintillation light from isolated segments is collected by optical fibers and registered with silicon photo-multipliers (SiPM). 
    The work contains the result of laboratory testing of the detector prototype. The detector channels were calibrated, energy resolution for every channel was obtained. Moreover, we present an advanced integral data acquisition and analysis technique based on Bayesian statistics, which will allow operation even during SEP events with very large fluxes.
    
    The work is motivated by the need for better measurement tools to study acceleration and transport of SEP in the heliosphere as well as by the need for the monitoring tool to mitigate radiation hazard for equipment and people in space.
}

\begin{document}

\maketitle

\section{Introduction}

The Sun is a magnetically active star. As a result of the processes of transformation of free magnetic energy occurring in active regions in the solar atmosphere, populations of solar energetic particles (SEPs) or solar cosmic rays (SCRs) with energies ranging from tens of keV to several GeV sporadically appear in the interplanetary (IP) medium (e.g., \cite{Malandraki2018,Miroshnichenko2015}). Due to the composition of the chemical elements of the Sun, the most numerous are populations of energetic electrons and protons. Among SEPs, there are also heavier nuclei from He to Fe, but in much smaller quantities.

According to the modern paradigm \cite{Reames2013,Malandraki2018}, the SEP events (SEPEs) are divided into two main groups: 1) impulsive events associated with coronal jets or impulsive solar flares, 2) gradual events with long-duration eruptive flares accompanied by fast ($v >\approx 1000$ km/s) coronal mass ejections (CMEs) and related large-scale shock waves. Usually impulsive SEPEs last several tens of minutes or a few hours and accelerate electrons predominately, whereas gradual SEPEs can last several days, manifesting themselves in the form of SEP radiation storms in the IP medium. Distinctive features of impulsive SEPEs are: fast-drifting type III radio bursts (associated with beams of accelerated electrons propagating through the plasma of the corona and IP medium), a relatively narrow (up to a few tens of degrees) cone of spread in the IP medium and good magnetic connection with a parent flare region on the Sun, an increased by $\sim 10^3 -10^4$ ratio of $^3$He/$^4$He, an increased by $\sim 10$ ratio of Fe/O ions relative to the nominal coronal values, the ionization state of Fe around 2. Gradual SEPEs are usually accompanied by slow-drifting type II radio bursts (associated with the plasma mechanism of radiation on a propagating shock wave), they are rich in protons, the average Fe/O ratio is of the order of 0.1, the ionization state of Fe is around 14, and wide particle spread in the heliographic longitudes (up to 180 degrees and even higher) and latitudes.

It is believed, within the framework of the modern paradigm, that the properties of impulsive SEPEs can be explained by the relatively faster (minutes) acceleration of charged particles in flare regions on the Sun at relatively low altitudes up to 50-100 Mm above the photosphere as a result of magnetic reconnection, whereas in gradual SEPEs particles can be accelerated for a long time (tens of hours) by a shock wave, moving away from the Sun and expanding in the IP medium. In periods of high solar activity, when events occur close in time to each other (within several hours), mixed or hybrid SEPEs also occur \cite{Kallenrode2003}.

Despite the widespread acceptance of the modern paradigm, the fundamental problem of SEPs/SCRs is still far from a complete solution. Some important questions remain. For example: how exactly are particles accelerated in impulsive and long-duration solar flares, what is the role of magnetic reconnection and eruptive magnetic flux ropes in particle acceleration and escape from the corona, where and how exactly do particles make longitude-latitudinal transport, how and where seed populations of particles are formed for further acceleration at shock waves. Moreover, the modern paradigm has an alternative. Some researchers believe that in all SEPEs, both impulsive and gradual ones, solar flares play a key role in the acceleration of electrons and protons, and coronal/IP shocks play only a secondary role (e.g., \cite{Miroshnichenko2018,Struminsky2019}). It is suggested that there are two phases of acceleration in solar flares. In the first phase, electrons are mainly accelerated to relatively low energies of $\sim 100$ keV, and in the second phase, simultaneous long-term acceleration of electrons and protons to relativistic energies occurs. The importance of simultaneous measurements of solar energetic electrons and protons in a wide range of energies for understanding the mechanisms of their acceleration is emphasized in \cite{Struminskii2020}. A reliable determination of the energy spectra of electrons and protons is important for determining their acceleration and transport mechanisms (e.g., \cite{Miroshnichenko2015}). Since electrons are lighter particles, they have much higher speeds at the same energies as protons. With a simultaneous injection, relativistic electrons arrive at a distant observer before protons. Based on this idea, a method of short-term (from a few to tens of minutes) prediction of the arrival of solar energetic protons by analyzing the arrival of energetic electrons was proposed in \cite{Posner2007}.

Besides the importance of studying SEPs from the point of view of solving the fundamental problem of SEPs/SCRs, the detection and investigation of SEPs are also of great practical importance. The “heavy” component of SEPs - protons and various ions up to Fe with energies above $\sim 50$ MeV - has the most serious negative effect on spacecraft, on its electronic components, and astronauts on board (e.g., \cite{Miroshnichenko2015,Malandraki2018}). Penetrating the magnetosphere, especially in high-latitude regions with a “quasi-open” geometry of magnetic field lines, SEPs can affect the Earth’s ionosphere and atmosphere, causing radio interference, interruptions in satellite orientation and navigation, as well as creating increased doses of radiation absorbed by pilots and passengers of aircrafts. Relativistic electrons capable of penetrating deep enough under the lining of spacecraft can cause the volume electrification effect associated with the formation of deep dielectric charging \cite{Baker1997,Lai2011}. If the fluxes of energetic electrons are large enough for a long time, then the charge does not have time to dissolve, reaches critical values, and causes micro-breakdowns, which lead to disruption of the on-board electronics.

SEPs with energies up to a few hundred MeV do not penetrate down to the Earth’s surface through the magnetosphere and atmosphere. They can be detected only with the help of instruments installed on board spacecraft. Many SEP instruments have been developed and launched into space (e.g., \cite{Wuest2007}). Of the latter, there are the Integrated Science Investigation of the Sun (ISIS) suit of detectors aboard Parker Solar Probe \cite{McComas2016} and the Energetic Particle Detector (EPD) instrument aboard Solar Orbiter \cite{Rodriguez-Pacheco2019}. Typically, such instruments are sets of charged particle telescopes - multilayer assemblies of semiconductor detectors. Sometimes separate thick scintillation detectors are added to the semiconductor detector system, the light signal from which is taken using photoelectronic multipliers (PMTs). These scintillation detectors are used as calorimeters, in which the bulk of the energy of the measured particle is released, as well as an anti-coincidence shielding. The use of multilayer scintillation detectors was previously unjustified in SEP telescopes due to the bulkiness of the PMTs used and the relatively poor energy resolution. However, the situation has changed in recent years, as it has become possible to replace bulky traditional PMTs with significantly more compact silicon photomultipliers - SiPMs (e.g., \cite{Buzhan2003,Dolgoshein2006}).

In this article, we present the results of the development and calibration of a laboratory prototype of a SEP detector-telescope (electrons with energies $\sim 1-10$ MeV, protons $\sim 10-100$ MeV), based on a multi-layer system of plastic scintillation detectors with compact SiPMs.

\section{Detector construction}

There are several requirements for the telescope. Firstly, it must have a good energy resolution (at least $\Delta E/E \approx 10-15$\%). Secondly, it should also work at high particle fluxes to measure the spectrum during intense SEPEs (up to $\sim 10^{6}-10^{7}$ particles per second). Thirdly, to use the detector in space, it must have small mass-dimensional characteristics. To optimize the detector geometry, we used Geant4 simulation. The results of the simulation are presented in \cite{Zelenyi2019}.

The optimal geometry of the detector strongly depends on the physics of radiation passing through matter. Firstly, the length of the scintillation detector is defined by the maximum penetration of particles of interest. For 10 MeV electrons and 100 MeV protons, the penetration through a scintillator is less than 7 cm. The final prototype length of 8 cm reasonably covers the energy region. The diameter of the detector is a more complicated problem. On the one hand, increasing the detector diameter increases the number of particles being captured and the maximum angle. On the other hand, larger diameter means heavier detector and worse energy resolution due to more wide particle angular distribution (this problem could be solved by additional transversal detector segmentation). For the prototype, we decided to use $3~cm$ diameter. Simulations show that such diameter allows us to capture the whole proton energy deposition and about 90\% of the electron avalanche for the electron energy of $7~MeV$ (for smaller energies the fraction of energy deposition is even larger). Those dimensions could be increased to to extend observation to more energetic particles or larger incident angles. It is also possible to add metallic layers to allow measurement of even higher energies without increasing the size of the detector (such technique is frequently used in accelerator detectors), but it is not possible to do so without increasing the detector mass.

The primary feature of the detector is its segmentation. A detector segment is a scintillator disk. The height of the disk could be varied to better measure specific energy region. The division makes it possible to obtain not only the total energy deposition of a particle but also its dependence on penetration depth. This information allows us to significantly increase the precision of energy reconstruction especially for protons, for which the shape of energy deposition has a well-known peculiarity - Bragg peak (Fig.~\ref{fig:bragg}). The analysis of the loss function shape also allows us to distinguish between electrons and protons with the same total energy deposition because they have a different per-segment deposition.

The height of the disk should balance two factors. On one hand, the thinner the detector segments are, the more accurate, in terms of discretization, the loss function is measured. But, on the other hand, the thinner the cylinder, the less light one will get from energy loss in it. The number of detected photons directly affects the energy resolution of a single segment. Also, there is a physical detection threshold for preamplifiers. Moreover, a lot of disks require more complicated and more massive electronics resulting in the device mass increase, power consumption, and telemetry volume. For the prototype, we decided to use disks of 0.4 cm height, which, according to simulations, provides a reasonable balance between the number of measurement channels and loss function measurement precision. To cover the length of 8 cm, we need 20 such disks, which also means 20 channels of measuring electronics.

Light from disks is transported via optical fiber to silicon photo-multiplier detectors (SiPM). Detectors are attached to front-end electronics with integrated preamplifiers and automatically adjusted power supply for bias voltage. The output of the front-end electronics is processed by a back-end amplifier and digitizer.

\section{Detector response}

\begin{figure}[t] 
    \centering
    \includegraphics[width=0.7\linewidth]{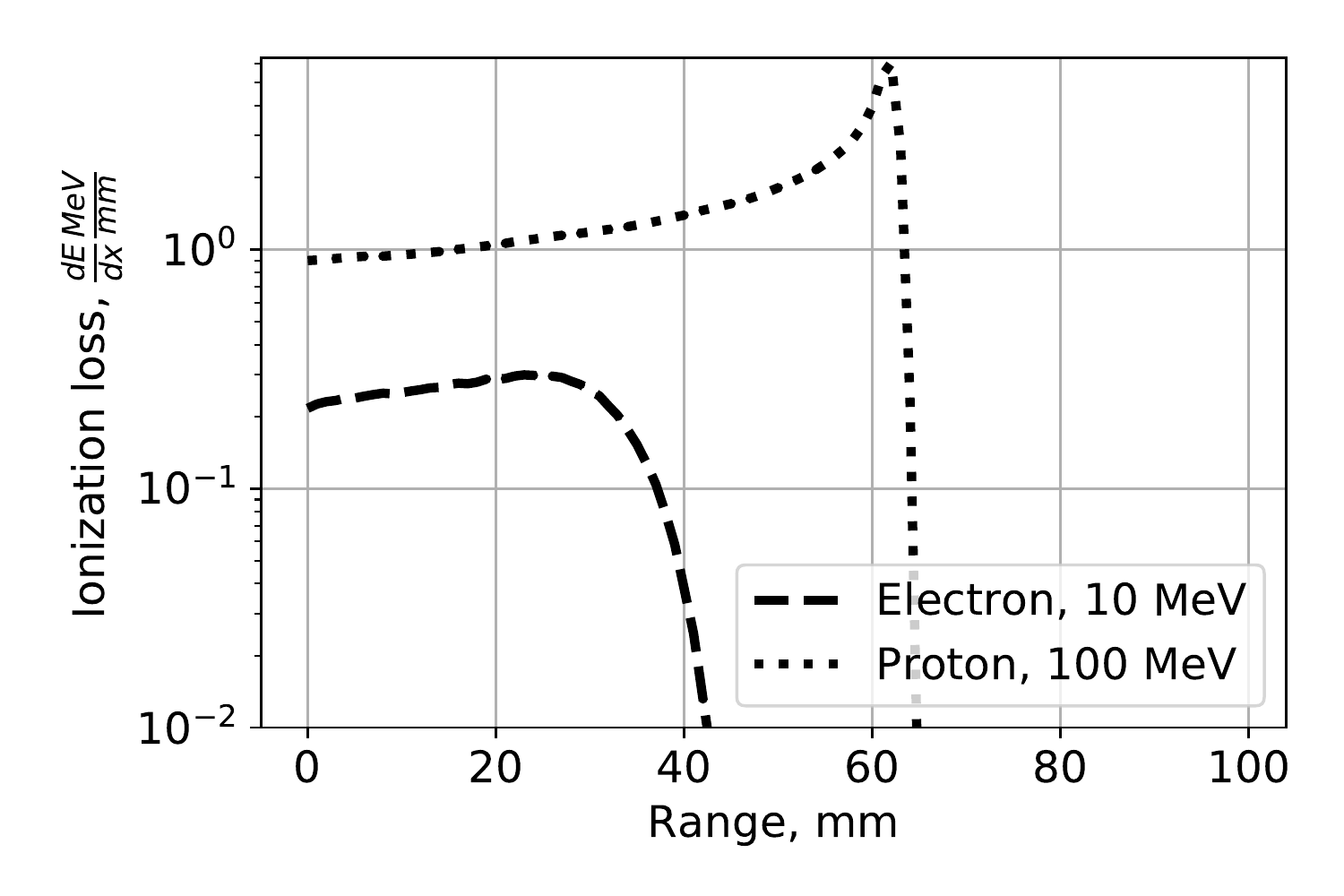}
    \caption{Proton and electron loss functions in a plastic scintillator. Protons have strongly marked Bragg peak.}
    \label{fig:bragg}
\end{figure}

A charged particle, when passing through a scintillator, loses energy, part of which is converted into visible light. Visible light is then detected by a photodetector. Charged particles have a characteristic function of energy loss per unit path length. For protons, for example, the loss function has a Bragg peak, the spatial position of which can be used to precisely determine the proton energy (Fig.~\ref{fig:bragg}). However, the electron loss function does not possess such a significant peak. Though with worse precision, electron energy is also resolvable from the curve. Moreover, the difference in proton and electron loss curve shapes allows one to distinguish these particles from each other.

By design, the detector should be able to operate in two modes:
\begin{itemize}
    \item Single-particle mode. This mode could be used when the total count rate is relatively low (up to $10^{5}$ particles per second). In this case, it is possible to analyze each particle separately. The energy and type of particle are restored from its loss curve by the method of maximum likelihood. 
    
    \item Integral mode. In this mode, the specific energy deposition of each particle is not measured, instead, the total light yield is integrated for each segment over a fixed time. Information on the spectrum of particles is reconstructed from the total curve of their losses obtained during the exposure. The spectrum is obtained by solving the inverse problem by the Turchin regularization method \cite{turchin}, or the least-squares method, or by fitting the spectrum of particles. 
\end{itemize}

\section{The prototype detector}

\begin{figure}[ht] 
    \centering
    \includegraphics[width=0.7\linewidth]{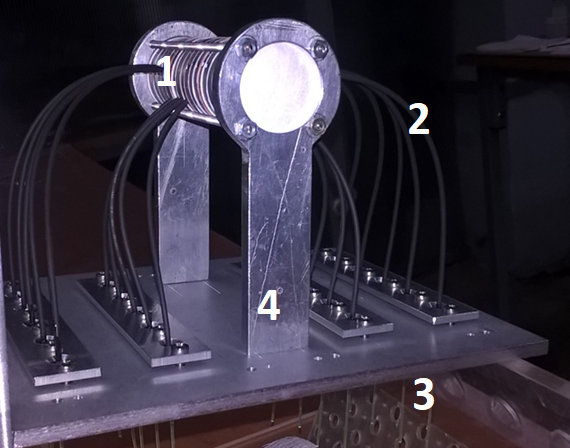}
    \caption{Picture of the detector prototype with the aluminum support structure. 1 - scintillator disks, 2 - optical fibers, wrapped in a black insulating coating, 3 - electronics with SIPMs attaching point, 4 - aluminium stand.}
    \label{prototype}
\end{figure}

\begin{figure}[ht] 
    \centering
    \includegraphics[width=0.45\linewidth]{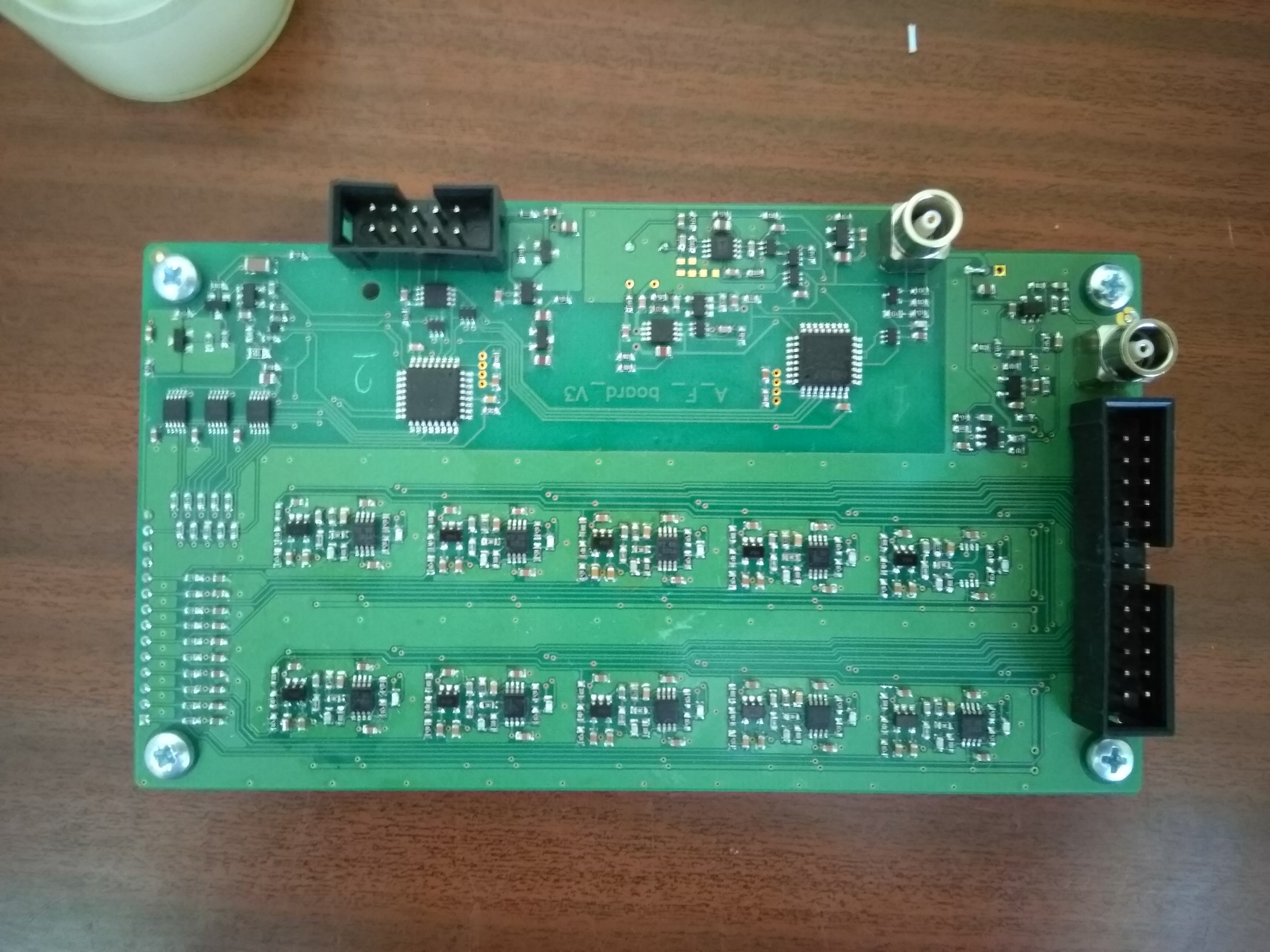}
    ~
    \includegraphics[width=0.45\linewidth]{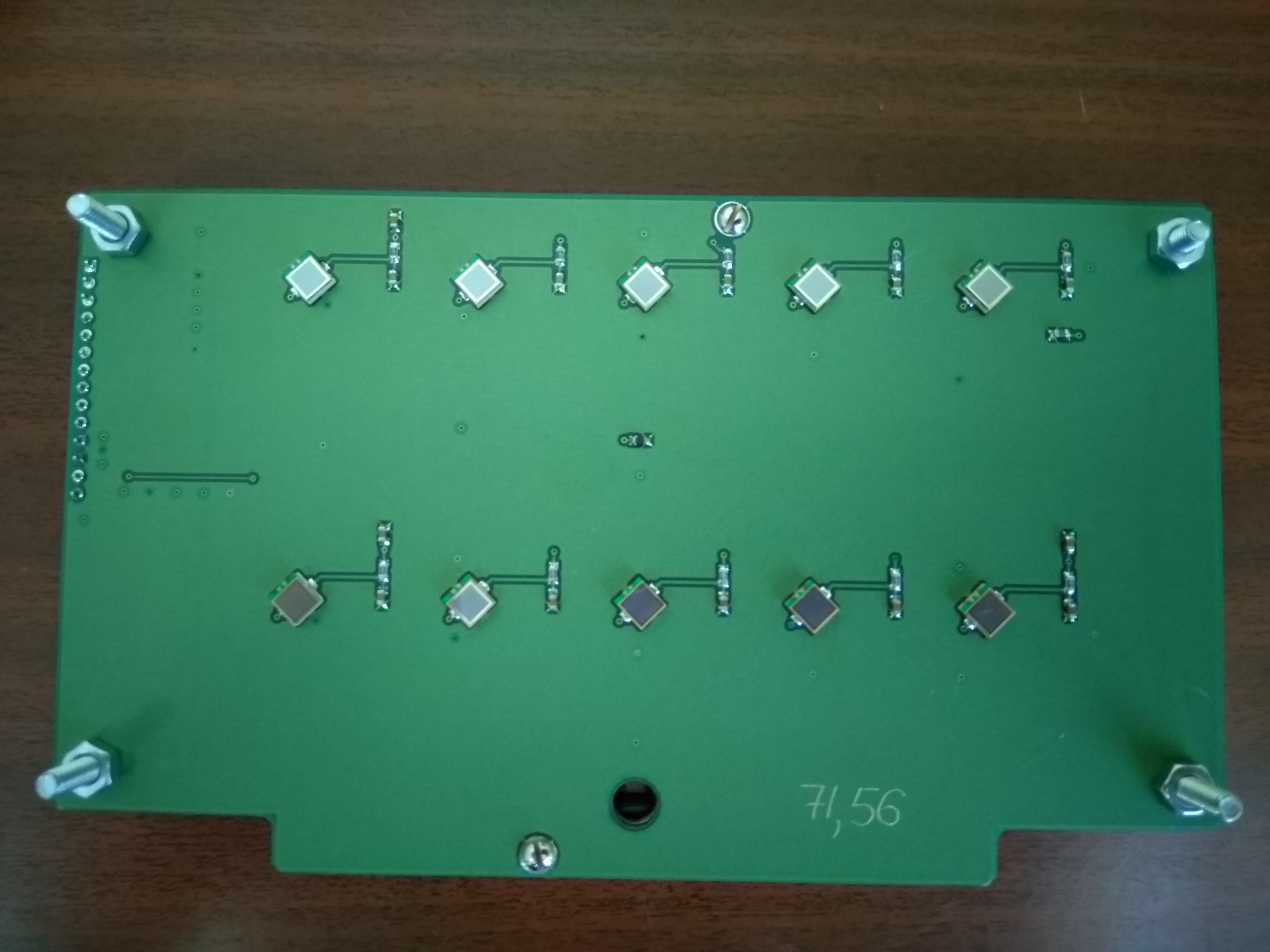}
    \caption{Picture of the detector electronic board with 10 SiPMs installed.}
    \label{electronics}
\end{figure}

The experimental prototype of the detector developed in INR RAS is a cylinder composed of 20 scintillation disks (Fig.~\ref{prototype}, Fig.~\ref{electronics}). A polystyrene-based material is used as a scintillator. It is possible to use different materials, but they have similar qualities, so for the prototype, we use the one that was easily accessible and widely used in detector construction. The diameter of a disk is 3 cm, the thickness is 0.4 cm. Each scintillator is wrapped in a reflective material - Tyvek, produced by DuPont company (\cite{tyvek}). It prevents optical cross-talk between channels and increases photon detection efficiency. The usage of Tyvek is a common practice in scintillator detector construction, its properties were studied for example in \cite{Abreu:2018ajc,ArteagaVelazquez:2005fj}. Scintillation light is detected using the Hamamatsu S12575-015P SiPM photodetectors \cite{hamamatsu}.

\begin{figure}[ht] 
    \centering
    \includegraphics[width=0.45\linewidth]{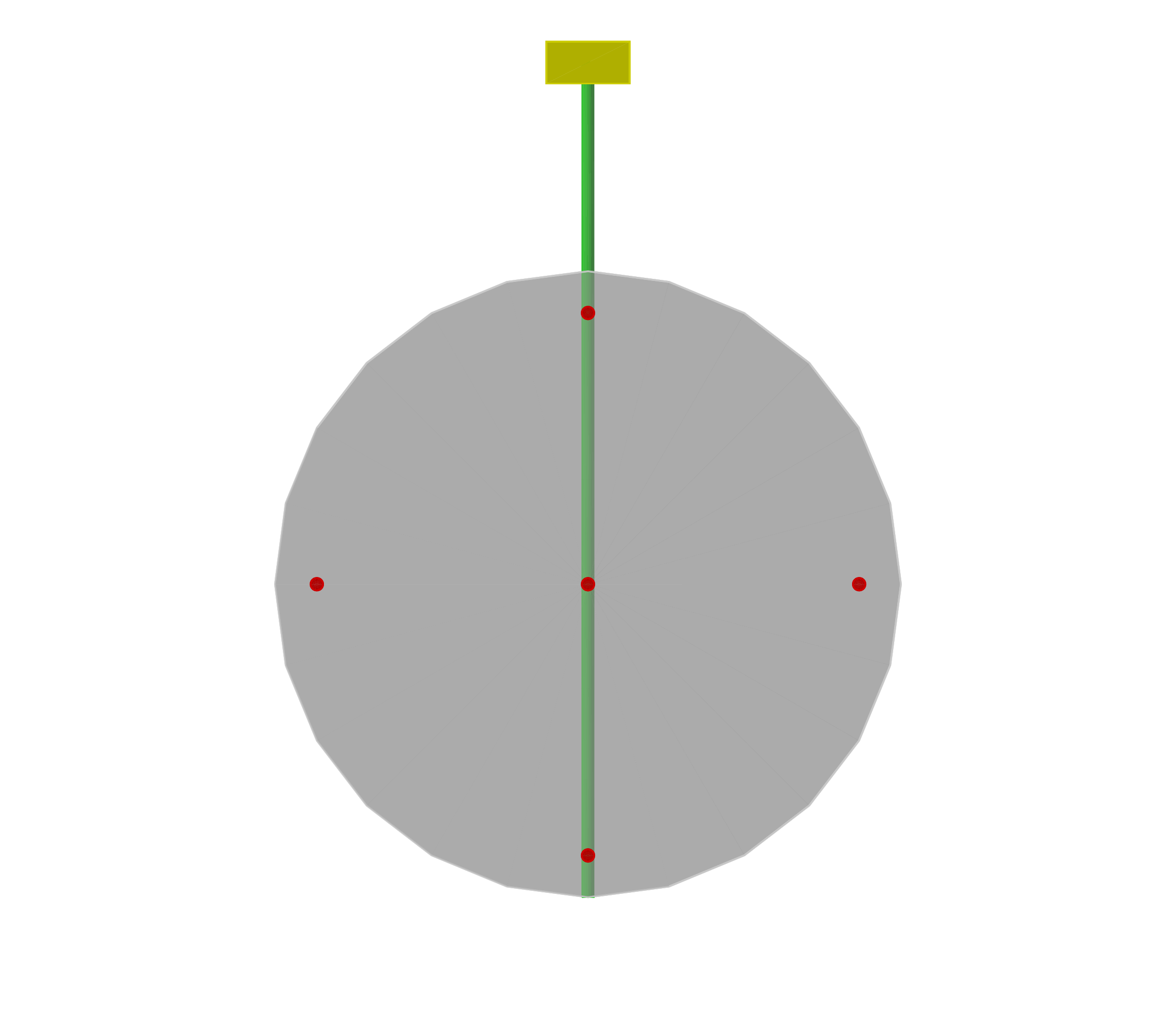}
    ~
    \includegraphics[width=0.45\linewidth]{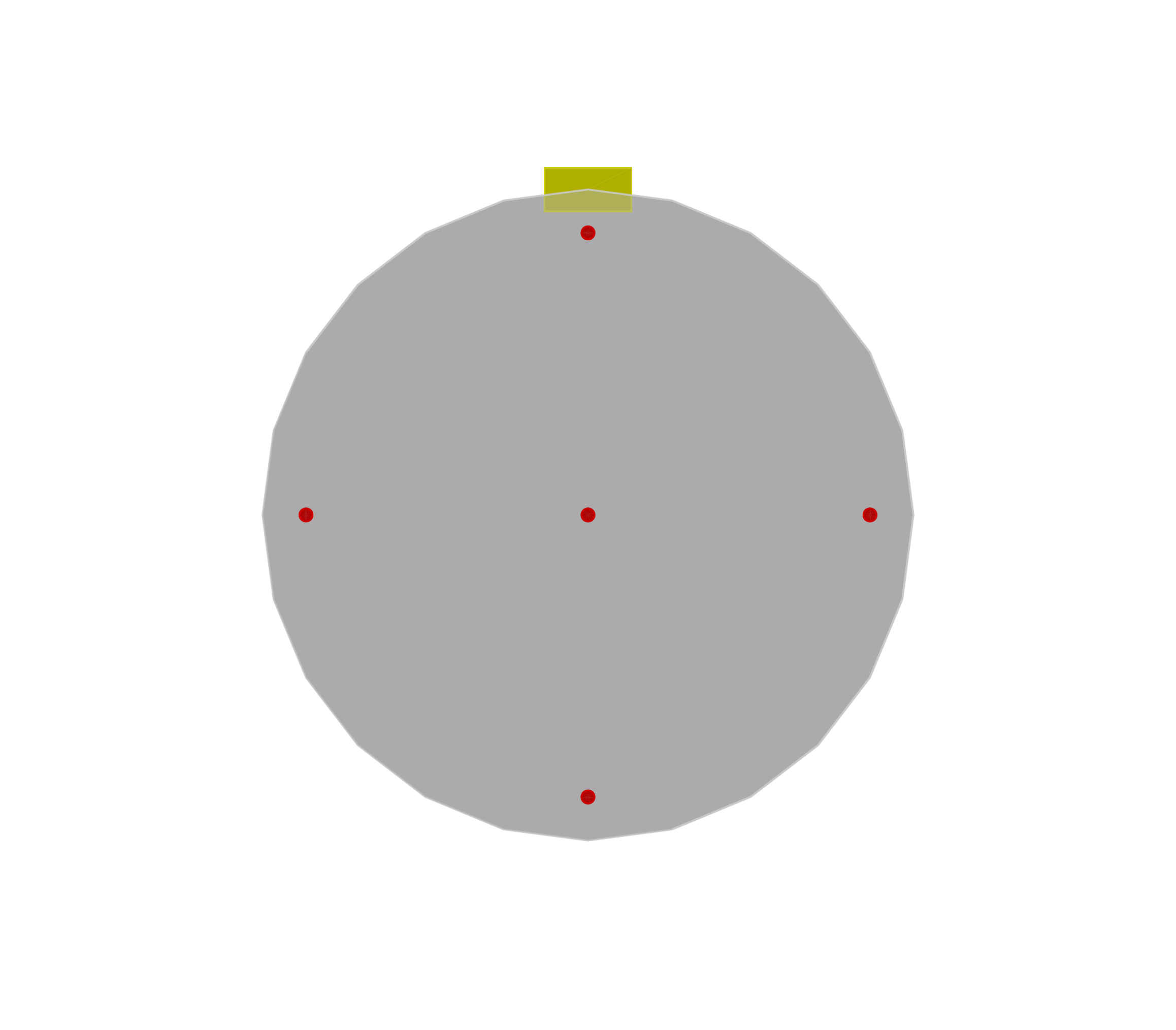}
    \caption{Two ways to collect photons from a scintillator disk. On the left picture - light collection via an optical fiber \ref{disks}. On the right picture - direct attachment of a SiPM to the scintillator disk. }
    \label{measurement_scheme}
\end{figure}

Two ways to register light with the SiPM were considered: 
\begin{itemize}
    \item to attach the photodetector directly to the disk via dry contact with special spring and/or optical glue;
    \item to collect scintillation light using optical fiber glued inside a groove in the disk and connect the fiber to an external photodetector.
\end{itemize}
A drawing of the operation scheme of both approaches is shown in Fig.~\ref{measurement_scheme}. 

\begin{figure}[ht] 
    \centering
    \includegraphics[width=0.6\linewidth]{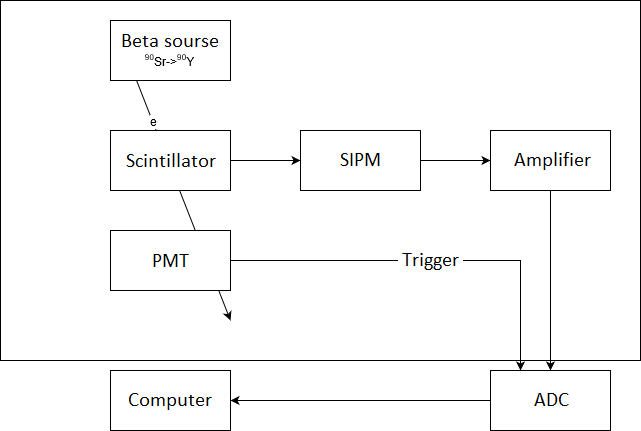}
    \caption{The scheme of a circuit being used for number of photoelectrons measurement.}
    \label{circuit}
\end{figure}

The decision about the measurement scheme was done basing on the experimental study of both approaches. The disks were irradiated with the $^{90}Sr$ laboratory beta source through a narrow collimator. It emits beta electrons with energies up to 1.5 MeV. Received signals were digitized and transmitted to a computer using the circuit shown in Fig.~\ref{circuit}. Two parameters were under consideration: the maximum number of photo-electrons and the uniformity of light collection. The number of photo-electrons is a parameter that shows the magnitude of the signal from the segment disk per unit of energy released within the disk by a high-energy particle. The uniformity of the light collection shows how strongly the signal of the detector depends on the point at which the particle enters it. 

Fig.~\ref{fig:photoelectrons} shows the results of the experiment. The picture on the left shows the points, to which the source was attached, the table to the right shows the resulting number of photoelectrons after calibration. It turned out that a design with a photodetector attached directly to the disk gives a 3 times larger number of photoelectrons compared to the design with optical fiber. However, the second method of attaching the photodetector provides almost complete uniformity of light collection in contrast to the first one. For example, when the SiPM is attached directly to the disk, the signal from a particle that hits the edge of the disk drops dropped by 30$\%$ comparing to the central hit. In the design with optical fiber, the difference between the same signals is $10-20\%$ (Table~\ref{tab:photoelectrons}). Since the difference in the number of photoelectrons by 3 times reduces the error only to the root of 3 times and non-uniformity could significantly complicate the analysis, the second design was chosen.

\begin{table}[ht]
\begin{minipage}{0.5\linewidth}
    \begin{tabular}{ | p{40pt} | p{80pt} | p{80pt} | }
        \hline
        $\beta$-source position & Direct attachment, photoelectrons & Optical fiber, 
        
        photoelectrons \\ \hline
        1 & 33 & 10 \\ \hline
        2 & 34 & 11 \\ \hline
        3 & 26 & 9 \\ \hline
        4 & 23 & 9 \\ \hline
        5 & 24 & 10 \\
        \hline
    \end{tabular}
    \caption{
        A uniformity of photo-collection test. The table shows the number of collected photoelectrons depending on the position of the $\beta$-source (Fig. \ref{fig:photoelectrons}) and attachment type.
    }
    \label{tab:photoelectrons}
\end{minipage}
~
\begin{minipage}{0.5\linewidth}
    \centering
    \includegraphics[width=0.65\linewidth]{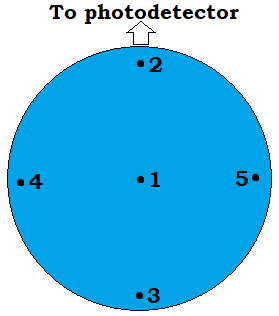}
    \captionof{figure}{$\beta$-source positions for the Table~\ref{tab:photoelectrons}.}
    \label{fig:photoelectrons}
\end{minipage}
\end{table}

The design with optical fibers has another significant advantage: the possibility to transport light far enough from the detector cylinder so one could provide additional radiation shielding for SiPMs and the electronics.
 
\begin{figure}[ht] 
    \centering
    \includegraphics[width=0.7\linewidth]{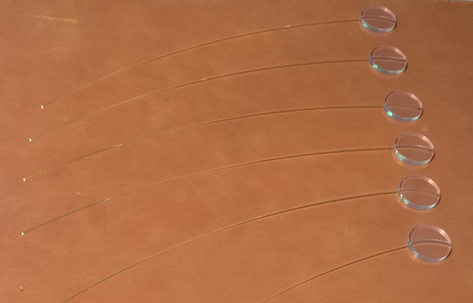}
    \caption{
        Picture of scintillator disks, which were used for the prototype. Disks are presented without a tyvek wrapper.
    }
    \label{disks}
\end{figure}

Fig.~\ref{disks} shows manufactured disks with grooves and glued fibers. Fig.~\ref{prototype} shows a picture of the assembled detector. A picture of one of the electronic boards with 10 SiPMs installed is shown in Fig.~\ref{electronics}. Two identical electronic boards (previously used in hadron calorimeters on accelerator experiments) are attached to the bottom of the breadboard so that optical fibers connect directly to the SiPMs. All SiPMs are located on a heat-conducting substrate, which makes it possible to equalize their temperature. There is a thermocouple on the electronic board that allows us to determine the temperature of the photodetectors. The slow control unit is connected to the electronics, supplying prescribed voltage to the SiPM taking into account its dependence on the temperature.

\begin{figure}[t] 
    \centering
    \includegraphics[width=0.7\linewidth]{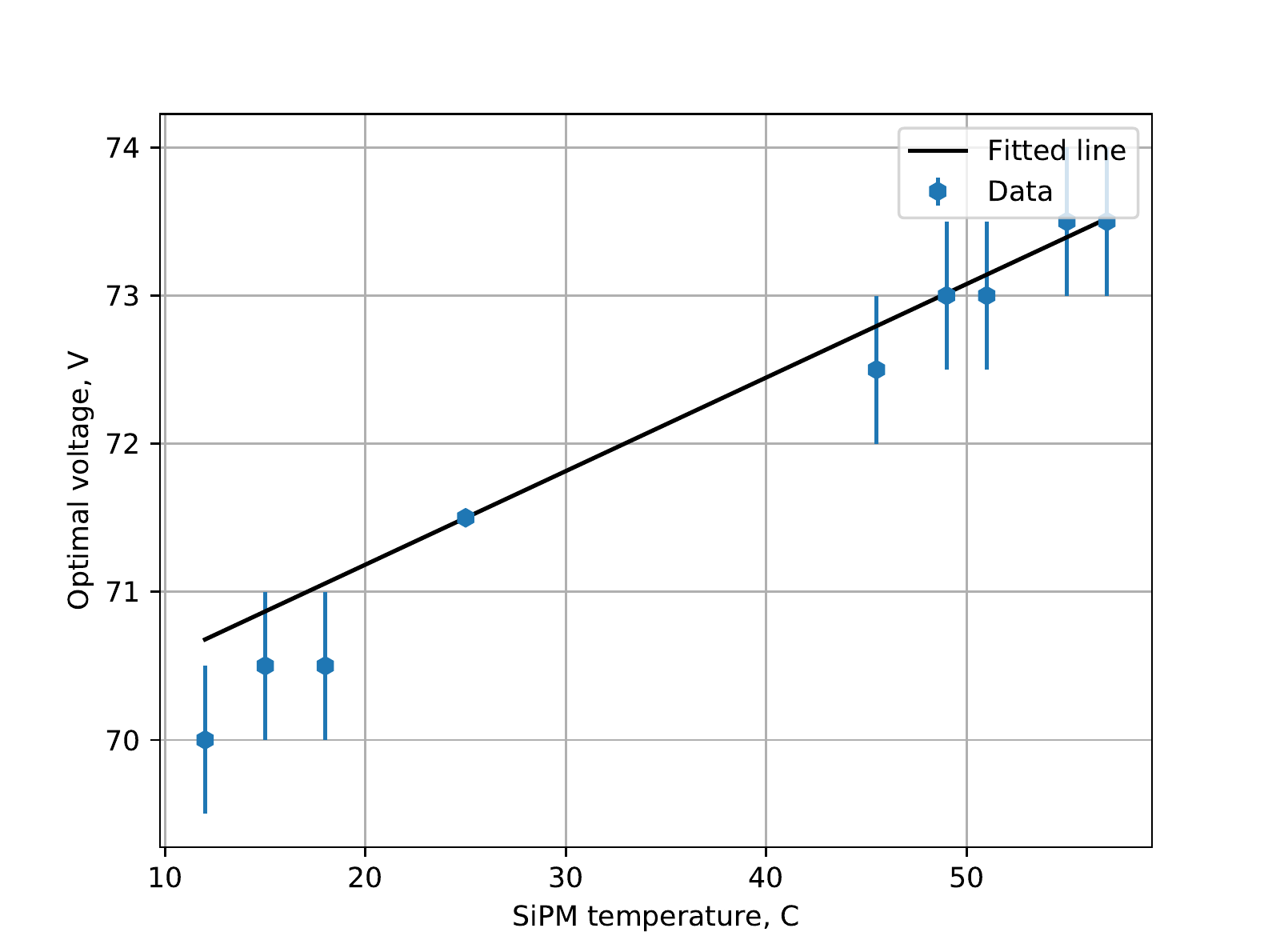}
    \caption{The result of temperature dependence test. The voltage required to obtain the same signal amplitude depending on the temperature.}
    \label{temperature}
\end{figure}

Setting SiPM voltage in accordance with temperature is crucial for detector designed to operate in an environment with large temperature differences. For example, the amplitude per one photoelectron increases by 2 times when the temperature drops from 25 $^\circ \mathrm{C}$ to 0 $^\circ \mathrm{C}$. The temperature dependence of the optimal voltage to be applied to a SiPM S12575-015P is well-known and described in the documentation \cite{hamamatsu}. However, in this work, it was verified experimentally. The disk and photodetector system were placed in a thermostat. The disk was irradiated using a laboratory beta source. Firstly, at the room temperature, the optimum voltage, specified in the documentation, was set on a SiPM. The optimal voltage for the room temperature was chosen as 71.5 V. The response of the system to beta radiation was measured. Then, at different temperatures, a voltage was selected so that the response of the disk to the $\beta$-source was the same as at the room temperature. The temperature dependence of the optimal voltage on the SiPM is shown in Fig.~\ref{temperature}. With an increase in SiPM's temperature of 1 degree Celsius, it is necessary to increase the voltage by $58.5~mV$. It is consistent with the dependence indicated in the documentation of the photodetector, which says that the voltage must be changed to $60~mV$ when the temperature changes by 1 degree.

\section{Detector calibration}

\begin{figure}[ht] 
    \begin{minipage}{0.5\linewidth}
        \centering
        \includegraphics[width=\linewidth]{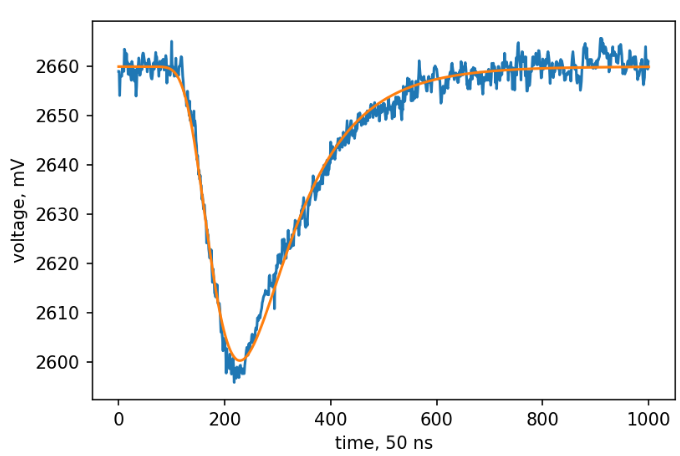}
        \caption{A single event signal fitting with the Landau-like distribution.}
        \label{psp_1}
    \end{minipage}
~
    \begin{minipage}{0.5\linewidth}
        \centering
        \includegraphics[width=\linewidth]{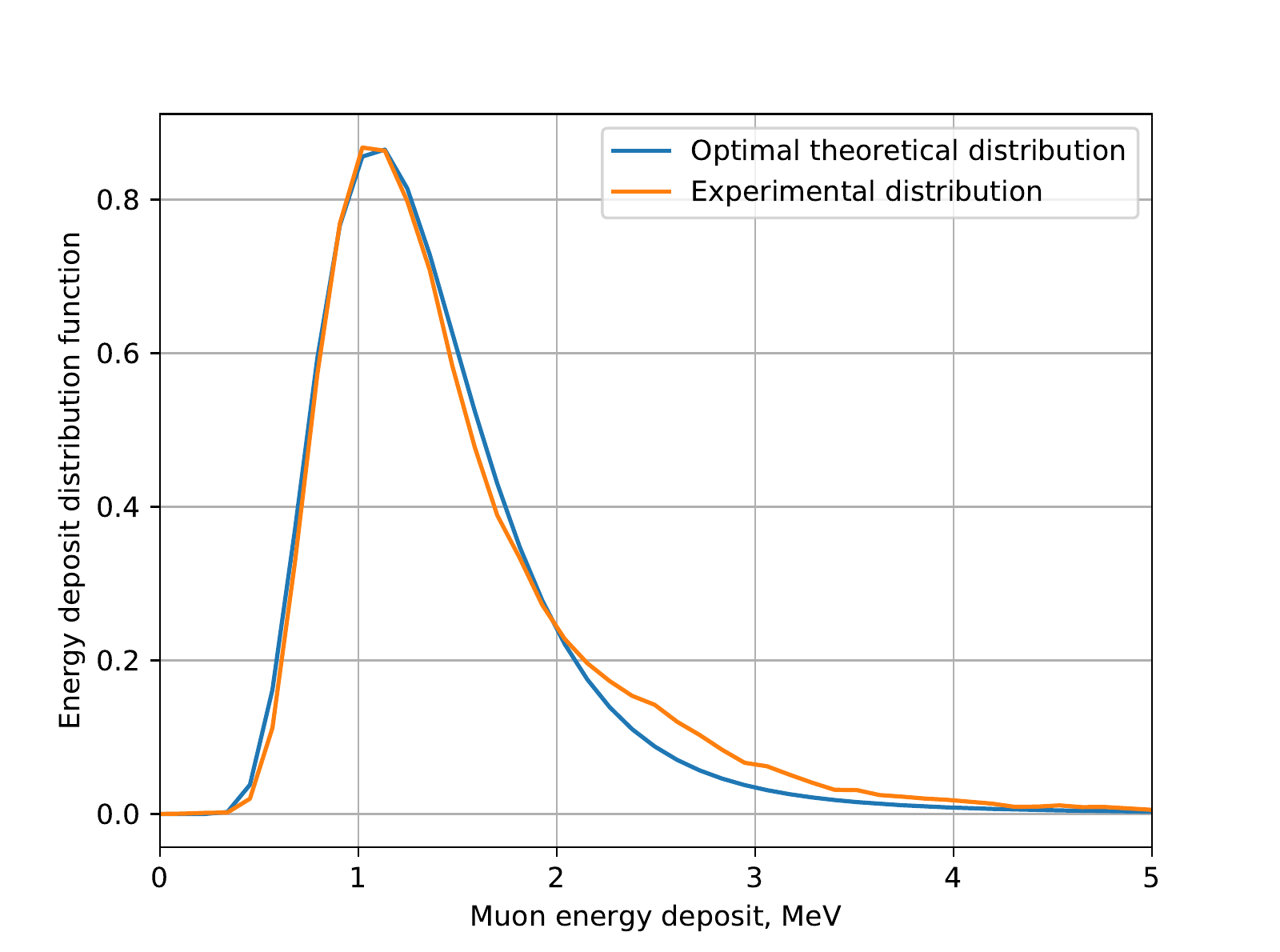}
        \caption{Muon signal amplitude distribution fit.}
        \label{distribution_fit}
    \end{minipage}
\end{figure}

Scintillator detectors are usually calibrated with atmospheric muons. Those muons, being minimally ionizing particles have a well-known energy deposition in the material. The average deposition is the same for all segments since the muon passes the whole detector with small relative energy loss. The typical muon signal from a measurement channel is shown in Fig.~\ref{psp_1}. To reduce the effect of random noise, we've fitted the signal with a Landau-like distribution ($e^{-(x + e^{-x})}$).

The distribution of energy deposition for each measurement channel (a disk + fiber + SiPM + pre-amplifier) was fitted with the expected muon signal distribution. The theoretical distribution was obtained with the following assumptions:

\begin{enumerate}
    \item The distribution of muons by the angle of incidence on the scintillator disk is $f(\theta) = \cos^2\theta$ with $\theta \in \left[ 0, \frac{\pi}{2} \right]$
    \item Muons always fly through the top and the bottom of the disk or through both sides of the disk.
    \item Error for the fixed signal has a normal distribution, but sigma depends on the value of the signal.
    \item Disk light collection heterogeneity is ignored.
\end{enumerate}

The obtained distribution formula looks as follows (the plot of the distribution is shown on figure \ref{distribution_fit}):

\begin{displaymath}
f(\varepsilon) = \left\{ 
\begin{array}{ll}
    f_0 \frac{\varepsilon_0^3}{\varepsilon^4 \sqrt{1 - (\frac{\varepsilon_0}{\varepsilon})^2}} & \textrm{$\varepsilon <$ 7.65 MeV}\\
    f_0 \varepsilon_0 \left(\frac{2r}{\varepsilon d}\right)^2 \sqrt{1 - \left(\frac{2r\varepsilon_0}{\varepsilon d}\right)^2} & \textrm{$\varepsilon \in (7.65~ \mathrm{MeV}, 7.72~\mathrm{MeV})$}
    \end{array} \right.
\end{displaymath}

Here $\varepsilon_0 = 1.02$ MeV, $r$ is disk radius, $d$ is disk thickness and $f_0$ is the normalization coefficient:

\begin{multline*}
    f_0^{-1} = \frac{1}{4}\sin\left(2\arccos(\frac{\varepsilon_0}{\varepsilon})\right)\big|_{\varepsilon_0}^{\varepsilon_{c}} + \frac{\arccos(\frac{\varepsilon_0}{\varepsilon})}{2}\big|_{\varepsilon_0}^{\varepsilon_{c}} + \\ \frac{r}{d}\arccos(\frac{\varepsilon_0}{\varepsilon})\big|_{\varepsilon(\frac{\pi}{2})}^{\varepsilon_{c}} - \frac{r}{2d}\sin\left(2\arccos(\frac{\varepsilon_0}{\varepsilon})\right)\big|_{\varepsilon(\frac{\pi}{2})}^{\varepsilon_{c}}
\end{multline*}

The detailed derivation of this formula is presented in the Appendix.

From theoretical distribution, it is known that the distribution peak corresponds to vertical muon energy deposit $\varepsilon_0 = 1.02$ MeV. Therefore, the peak position grants the calibration ratio. Here we assume that the ratio between the amplitude and the energy deposition is linear. 

\begin{figure} 
    \begin{minipage}{0.5\linewidth}
        \centering
        \includegraphics[width=\linewidth]{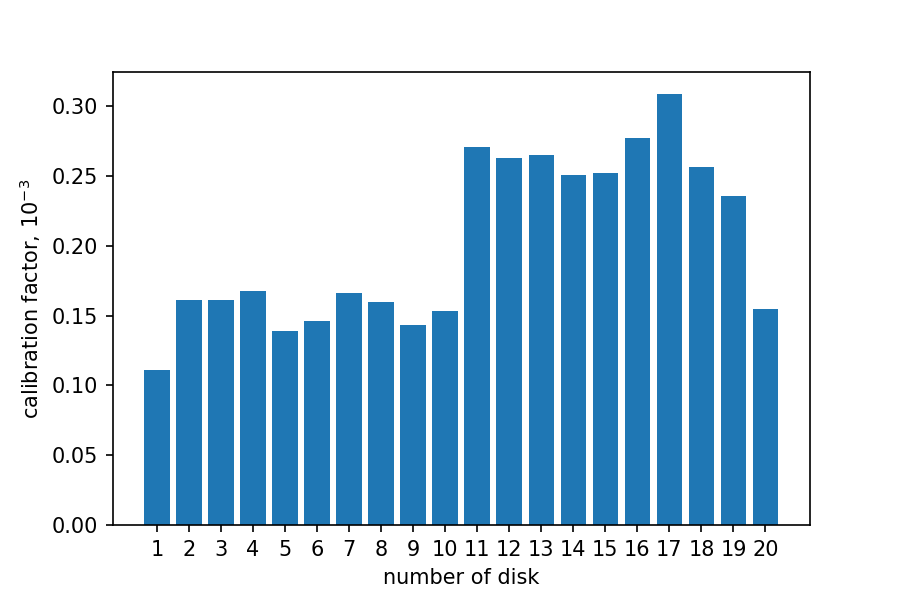}
        \caption{The calibration coefficients for all detector segments. }
        \label{calibration_results}
    \end{minipage}
    ~
        \begin{minipage}{0.5\linewidth}
        \centering
        \includegraphics[width=1\linewidth]{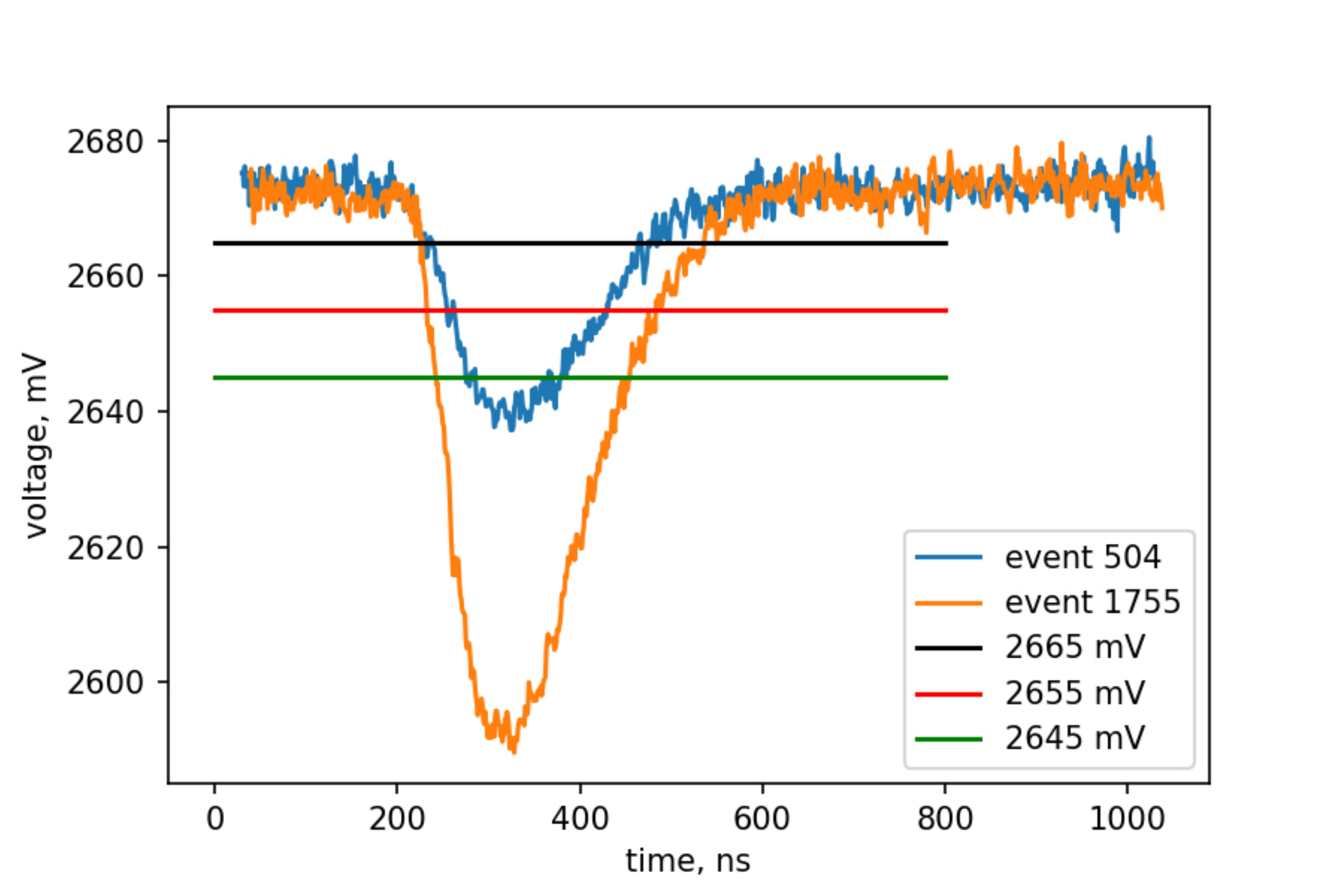}
        \caption{Threshold levels at a signal.}
        \label{lvls_signal}
    \end{minipage}

\end{figure}

The resulting conversion coefficients for different segments are presented in Fig.~\ref{calibration_results}. The difference in calibration coefficients between the first 10 channels and the last 10 channels appeared because they were connected to different readout boards. The resolution is, as expected, similar for all detectors and is about 30\% for 1~MeV deposition (as could be seen in Fig.~\ref{distribution_fit}). One must note that the relative resolution is inversely proportional to the square root of the deposited energy, so it will be better for protons compared to electrons (see Fig.~\ref{fig:bragg}). Also, the actual resolution of the whole detector is not limited by the resolution of a single segment but can be constrained much better by means of deposition shape analysis.

As a result, we obtained transformation coefficients from amplitude to the energy (individual for each measurement channel). The signal shape analysis could also be used to reduce the effect of signal pile-up for higher count rates (\cite{Chernov:2018ysc}).

\section{Width-based signal processing}

\begin{figure}[t]
    \begin{minipage}{0.5\linewidth}
        \centering
        \includegraphics[width=\linewidth]{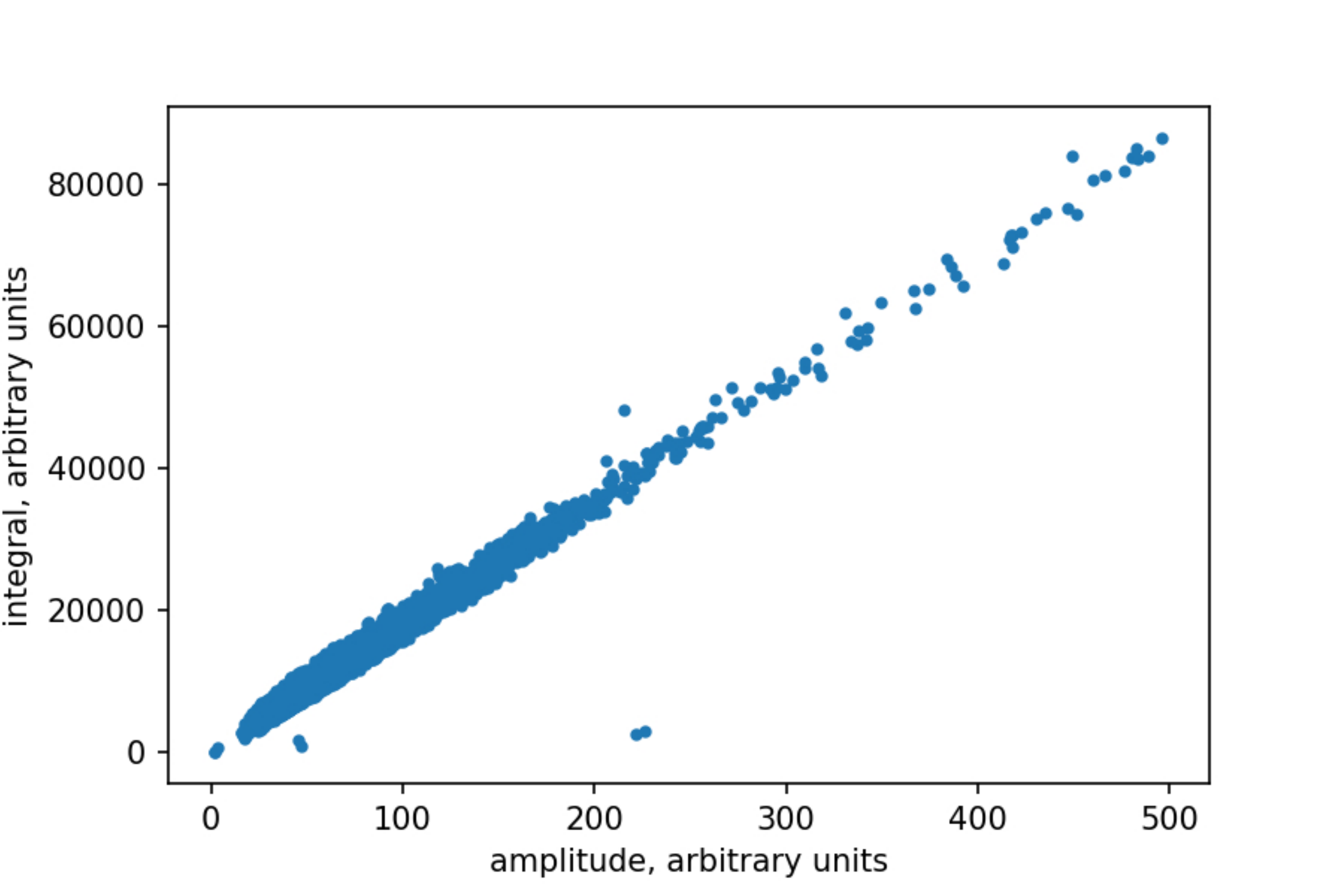}
        \caption{Relationship between integral and amplitude of a signal.}
        \label{fig:int_ampl}
    \end{minipage}
    ~
    \begin{minipage}{0.5\linewidth}
        \centering
        \includegraphics[width=\linewidth]{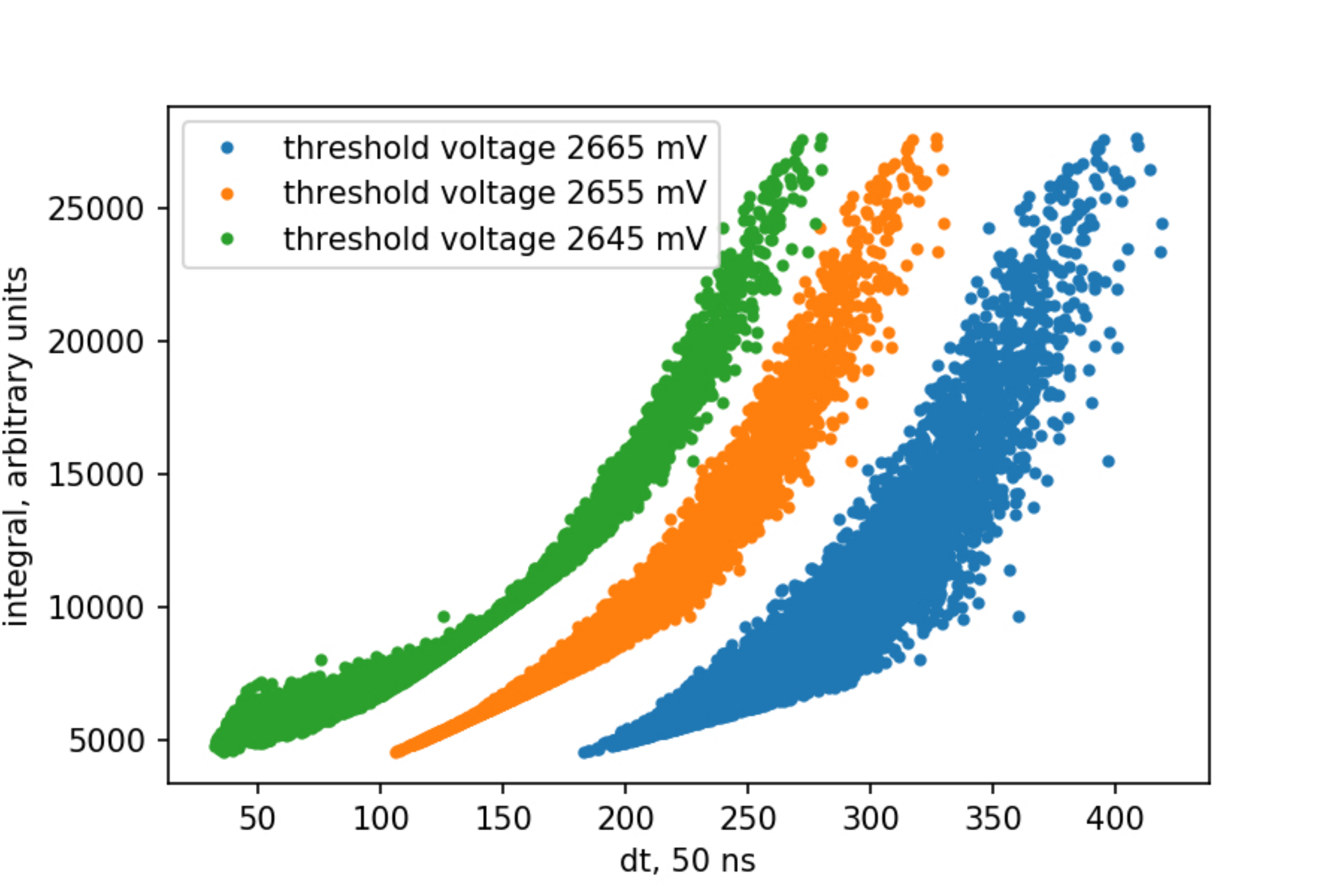}
        \caption{The relation between signal width and the amplitude for different threshold values.}
        \label{threshold_voltages}
    \end{minipage}
\end{figure}

Full signal shape analysis requires rather fast electronics and a lot of processing power, not always available on a spacecraft. So we considered different analysis procedure which relies on hardware-based peak detection instead of software-based one. The idea of the method is to register the time during which the signal exceeds the given threshold and then restore its amplitude based on a signal width instead of its height or integral (like it is shown in Fig.~\ref{lvls_signal}). The electronics is currently under development, but in this work, we present only some consideration based on the data obtained during calibration. As shown in Fig.~\ref{fig:int_ampl}, there is a linear relationship between the integral and the amplitude of signals.

\begin{figure}[ht]
    \centering
    \includegraphics[width=0.47\linewidth]{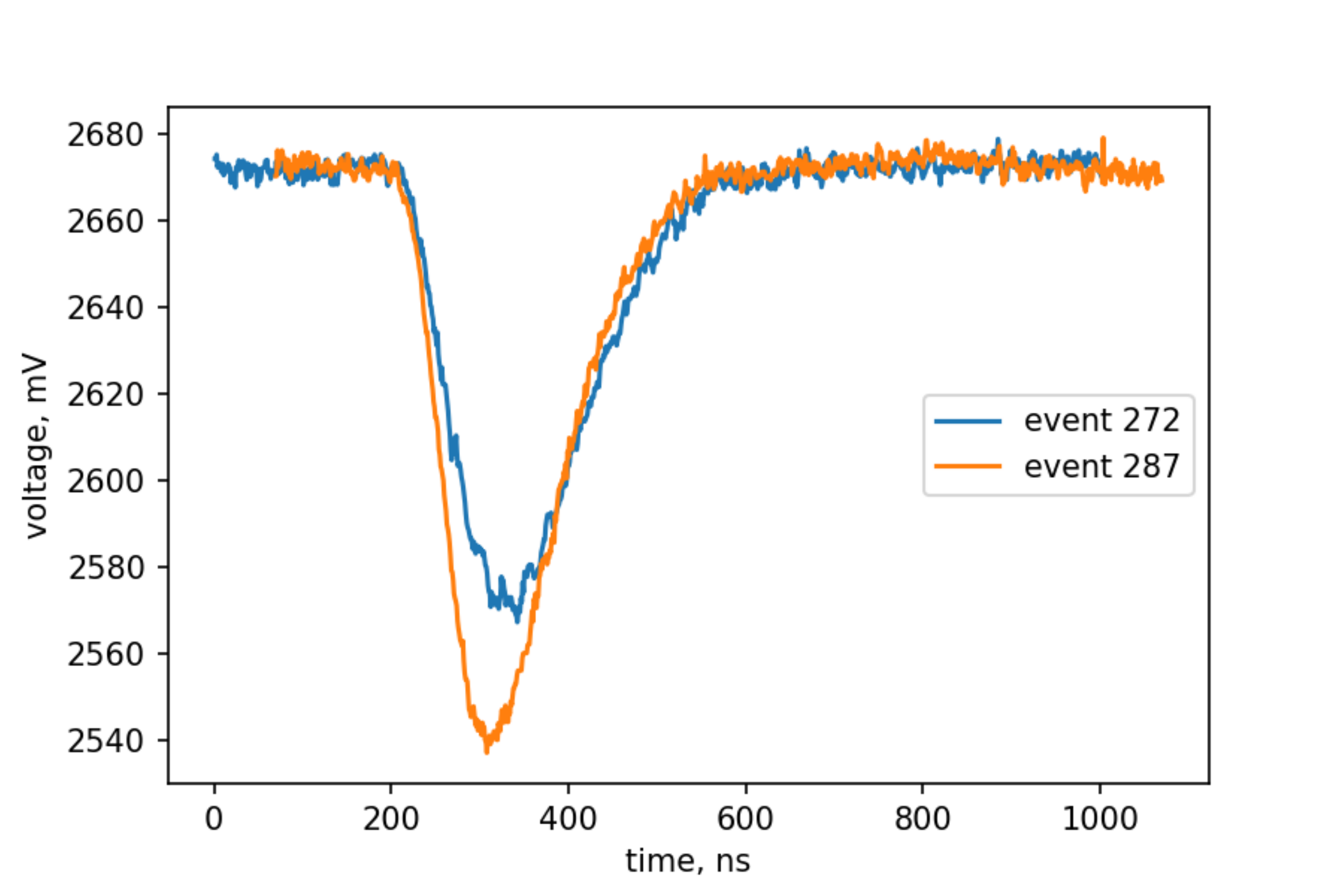}
    ~
    \includegraphics[width=0.475\linewidth]{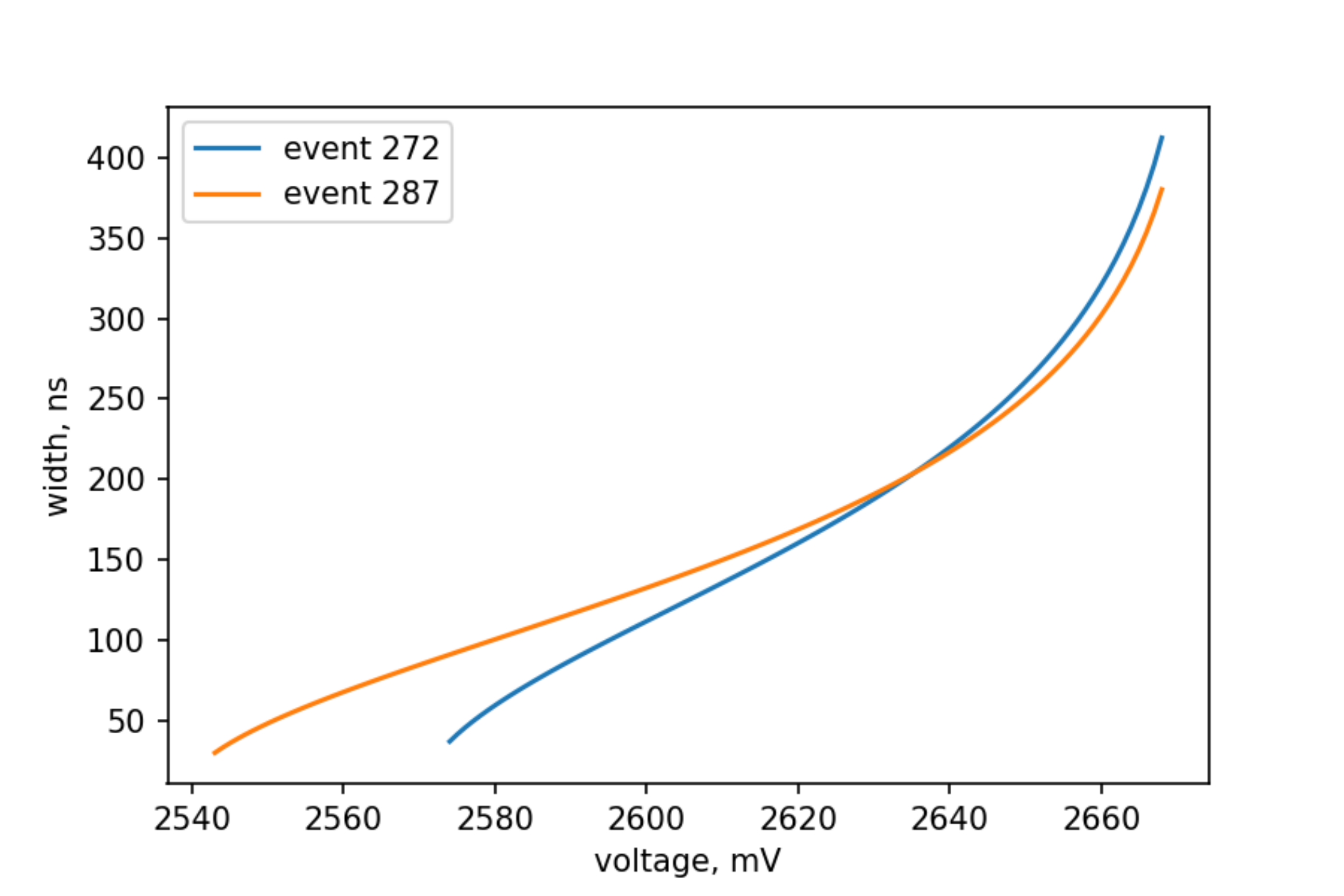}
    \caption{On the left the shape of two sample signals taken from calibration data. On the rite the signal width versus threshold value for those signals.}
    \label{intersection}
\end{figure}

The relation between the width with the given threshold and the signal amplitude is shown in Fig.~\ref{threshold_voltages}. It could be seen that while the dependence is not linear, it is possible to build a monotonic function to reconstruct the amplitude. Fig.~\ref{intersection} shows the signal shapes (on the left) and the relation between the threshold and resulting signal width for those signals (to the right). 

\begin{figure}[t]
    \begin{minipage}{0.5\linewidth}
        \centering
        \includegraphics[width = \linewidth]{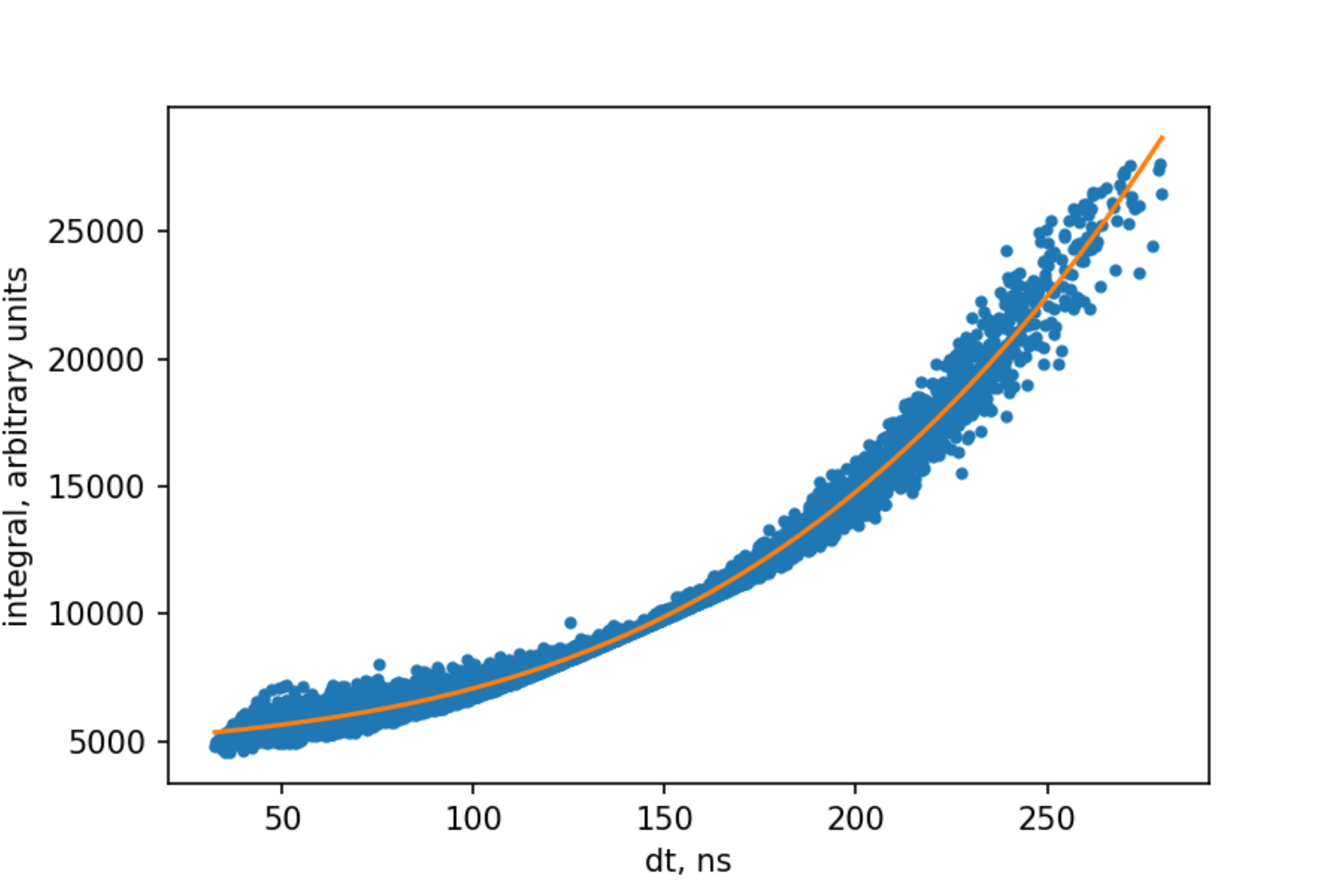}
        \caption{Amplitude reconstruction data for a threshold of 2645 mV}
        \label{fit_func}
    \end{minipage}
    ~
    \begin{minipage}{0.5\linewidth}
        \centering
        \includegraphics[width = \linewidth]{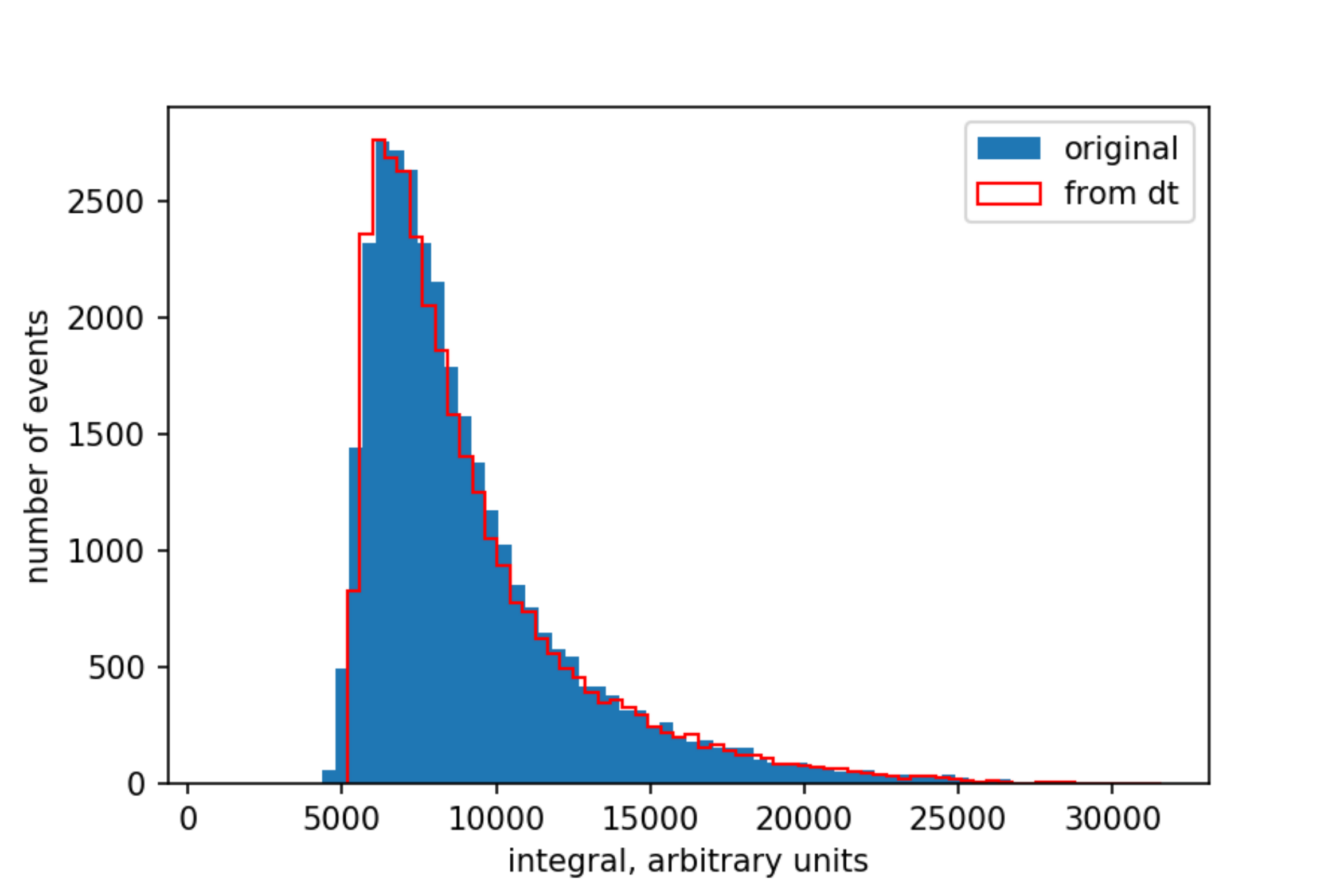}
        \caption{Comparison of distribution obtained via full signal processing (blue) and the distribution, obtained by signal width analysis (red).}
        \label{int_dist}
    \end{minipage}
\end{figure}

Based on given data, we chose a threshold value of 2645 mV because it proved to provide the best monotony with the smallest spread. The resulting distribution was fitted by the function $f(x) = a_0 + a_1x + a_3x^3$ (Fig.~\ref{fit_func}).

Finally, to verify that this approach recovers information about the signal, we compared the distribution of the amplitudes. Fig.~\ref{int_dist} shows both the original spectrum obtained from full signal analysis and one obtained from the signal width only. To get this picture, we performed partial blind analysis, meaning that we used only a small fraction of the data for calibration, and constructed the distributions for the whole data set. It could be seen that using only the signal width does no strongly affect resulting amplitude distribution. Thus, the calibrated detector with new electronics will be able to restore the spectrum of signals for particles of selected energies. 

The results from muons could not be simply extrapolated for protons and electrons since for that we need calibration data in a given region, but such calibration could be done either by using particles from proton accelerators, or by feeding generator pulses directly to the electronics board.

\section{Differential mode data analysis}

\begin{figure}[h]
    \centering
    \includegraphics[width=0.6\linewidth]{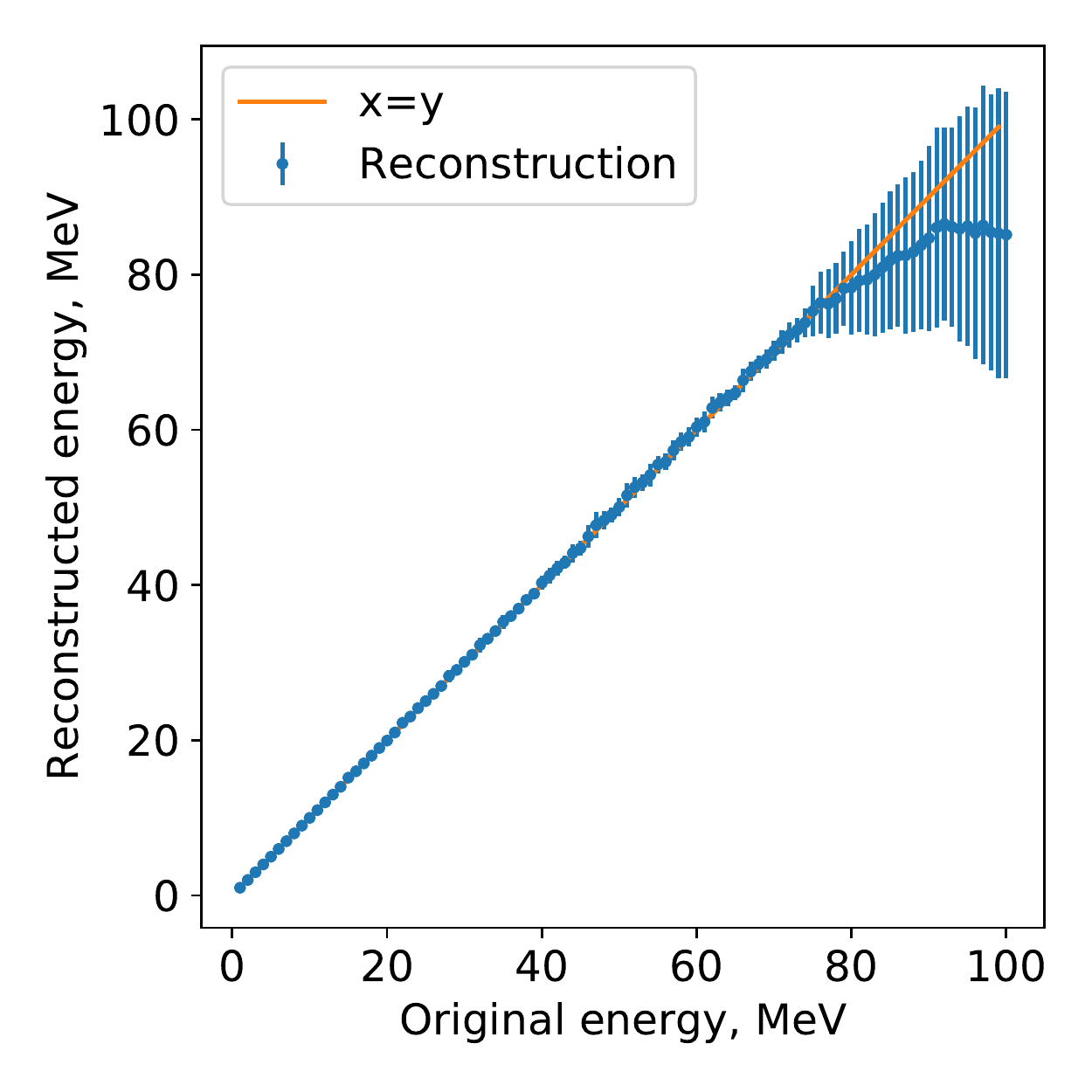}
    \caption{A reconstruction efficiency for protons with uniform angle distribution in a differential mode.}
    \label{fig:differential_mode_protons}
\end{figure}

A segmented detector provides a unique possibility to separate particles by type and identify their energy. The information about energy could be obtained both from the total number of photoelectrons - energy deposition in the whole detector and from the shape of the energy deposition. For protons and heavier ions Bragg peak (Fig.~\ref{fig:bragg}) allows acquiring very good (better than 5\% in most of the energy region) precision by simply comparing the experimental deposition shape with the simulation (like it is done in \cite{Zelenyi2019}). The protons could be separated from electrons and heavier ions by the energy deposition shape and the ratio between total energy deposition and the penetration length (penetration length is higher for a particle with the same energy, but lower mass). As one can see at Fig.~\ref{fig:bragg}, the ratio of energy depositions for the same penetration length is up to 5 times. The results of direct reconstruction for protons are presented at Fig.~\ref{fig:differential_mode_protons}. The drop in reconstruction efficiency for high energies is due to the losses of fractions of proton energy with large incident angles. It could be reduced by changing the detector size. For electrons, the energy reconstruction efficiency strongly depends on the detector position and specific detector case design which requires additional study. 

The limitation of the differential mode is the total count rate. When the number of particles exceeds electronics capabilities (ranging from $10^4$ to $10^7$ Hz depending on the specific electronics system), it is no longer possible to register individual particles. A further limitation is imposed by the on-board processing power required to evaluate all events.

\section{Integral mode data analysis}

\label{sec: Integral mode data analysis}

With the increase of particle flux, the detector reaches the maximum count rate for discrete data processing, therefore, the integral mode is required. In the integral mode, data is being collected over a fixed time, then particle energy spectrum is restored from scintillator energy deposition and previously collected detector calibration data.

Let $\vec{f}$ be the vector of measured energy deposition values from the first to the last disk. Let $\tilde K(E, y)$ be the energy deposition of a particle with initial energy $E$ MeV on the coordinate $y$ centimeters of the scintillator (can be approximated from calibration data). Let $\varphi(E)$ be the spectrum of incident cosmic particles. Thus, the problem can be  formulated as following: to find the spectrum, the integral equation (\ref{integral_equation}) should be solved.
\begin{equation}\label{integral_equation}
    f_i = \int_{E_{min}}^{E_{max}}\left(\int_{y_i}^{y_{i+1}} \tilde K(E, y)dy\right) \varphi(E) dE
\end{equation}
where $y_i$ - coordinate of the beginning of $i$-th disk.

Spectrum $\varphi(E)$ can be decomposed in basis $\{T_i(E)\}_{i=1}^I$:
\begin{equation}\label{phi_decomposition}
    \varphi(E) = \sum_{i=1}^I \varphi_i T_i(E)
\end{equation}
where $\varphi_i$ - $i$-th coefficient of spectrum decomposition.

Let $K_{i, n}$ be the energy deposition of particles with spectrum $T_i(E)$ in $n$-th disk.
\begin{equation}
    K_{i, n} = \int_{E_{min}}^{E_{max}} \left(\int_{y_n}^{y_{n+1}} \tilde K(E, y) dy \right) T_i(E) dE
\end{equation}
As a result, we have linear equation:
\begin{equation}\label{matrix_equation}
    \vec{f} = K \vec{\varphi}
\end{equation}

Both forms of the problem (\ref{integral_equation}) and (\ref{matrix_equation}) appear to be ill-posed, so they require special solution techniques. Some of those techniques are studied in \cite{SOHO-EPHIN}. One of the possible ways to do it is Turchin statistical regularisation \cite{turchin, turchin2}. The concept of this method is that in physical processes almost all spectra are relatively smooth. With additional information on function smoothness, one can solve equation \ref{matrix_equation} correctly and find particles spectrum $\varphi(E)$.

A solution of the equation \ref{matrix_equation} can be found as
\begin{equation}
    \vec{\varphi}_{opt} = E[\vec{\varphi} | \vec{f}] = \int \vec{\varphi} P(\vec{\varphi} | \vec{f}) d \vec{\varphi}
\end{equation}
By Bayes' theorem:
\begin{equation}
    P(\vec{\varphi} | \vec{f}) = \frac{P(\vec{\varphi})P(\vec{f} | \vec{\varphi})}{\int d \vec{\varphi} P(\vec{\varphi})P(\vec{f} | \vec{\varphi})}
\end{equation}
Every arbitrary distribution $P(\varphi)$ can be used as a prior information about $\varphi(E)$. For instance, we will use information that the $\varphi(E)$ function is relatively smooth. Let $\Omega$ be the matrix of average values of the second derivatives
\begin{equation}
    \Omega_{ij} = \int_{E_{min}}^{E_{max}}\frac{d^2 T_i(x)}{dx^2} \frac{d^2 T_j(x)}{dx^2} dx
\end{equation}
We express smoothness by parameter $\alpha$
\begin{equation}
    \alpha \int (\vec{\varphi}, \Omega \vec{\varphi})P(\vec{\varphi}) d \vec{\varphi} = 1
\end{equation}
This condition should bring as little information about the solution as possible, it means that information entropy has a minimum value
\begin{equation}
    \int P(\vec{\varphi})\ln{P(\vec{\varphi})} d \vec{\varphi} \rightarrow \min{}
\end{equation}
The solution to this problem is a parametric distribution:
\begin{equation}
    P(\vec{\varphi}) \sim \exp{-\frac{1}{2} (\vec{\varphi}, \alpha \Omega \vec{\varphi})}
\end{equation}
The regularization parameter $\alpha$ can be chosen arbitrarily.

Measured value $\vec{f}$ can have arbitrary distribution, but usually when measuring physical quantities, one has values that are distributed according to the normal distribution
$$P(\vec{f} | \varphi) = \frac{1}{(2\pi)^{N/2} \sqrt{\det{\Sigma}}}\exp{-\frac{1}{2}(\vec{f} - K \vec{\varphi})^T \Sigma^{-1} (\vec{f} - K \vec{\varphi})}$$
The product of the the prior distribution and the distribution of the measured quantity has the form of a multidimensional normal distribution. That allows to find $\vec{\varphi}$ and covariance matrix analytically
$$\vec{\varphi}_{opt} = (K^T \Sigma^{-1}K + \alpha \Omega)^{-1} K^T \Sigma^{-1T} \vec{f}$$
$$cov = (K^T \Sigma^{-1} K + \alpha \Omega)^{-1}$$

If the quantity $\vec{f} $ has a distribution other than normal, then the analytical solution is not always easy to find. In that case, we should find mode, expectation, and dispersion of the distribution using Monte-Carlo techniques.

Knowing the coefficients of the basis and the covariance matrix of the vector, we can find the spectrum functions $\varphi(E)$ at any point lying between $E_ {min}$ and $ E_{max}$:
$$\varphi(E) = \sum_{m=1}^M \varphi_m T_m(E)$$
and the error in calculating the spectrum at this point:
$$\delta \varphi(E) = \sqrt{\vec{T}^T(E) \cdot cov \cdot \vec{T}(E)},$$
where $\vec{T}(E)$ - vector of values of basis functions at the point $E$.

\begin{figure}[h]
    \begin{minipage}{0.5\linewidth}
        \center{\includegraphics[width=\textwidth]{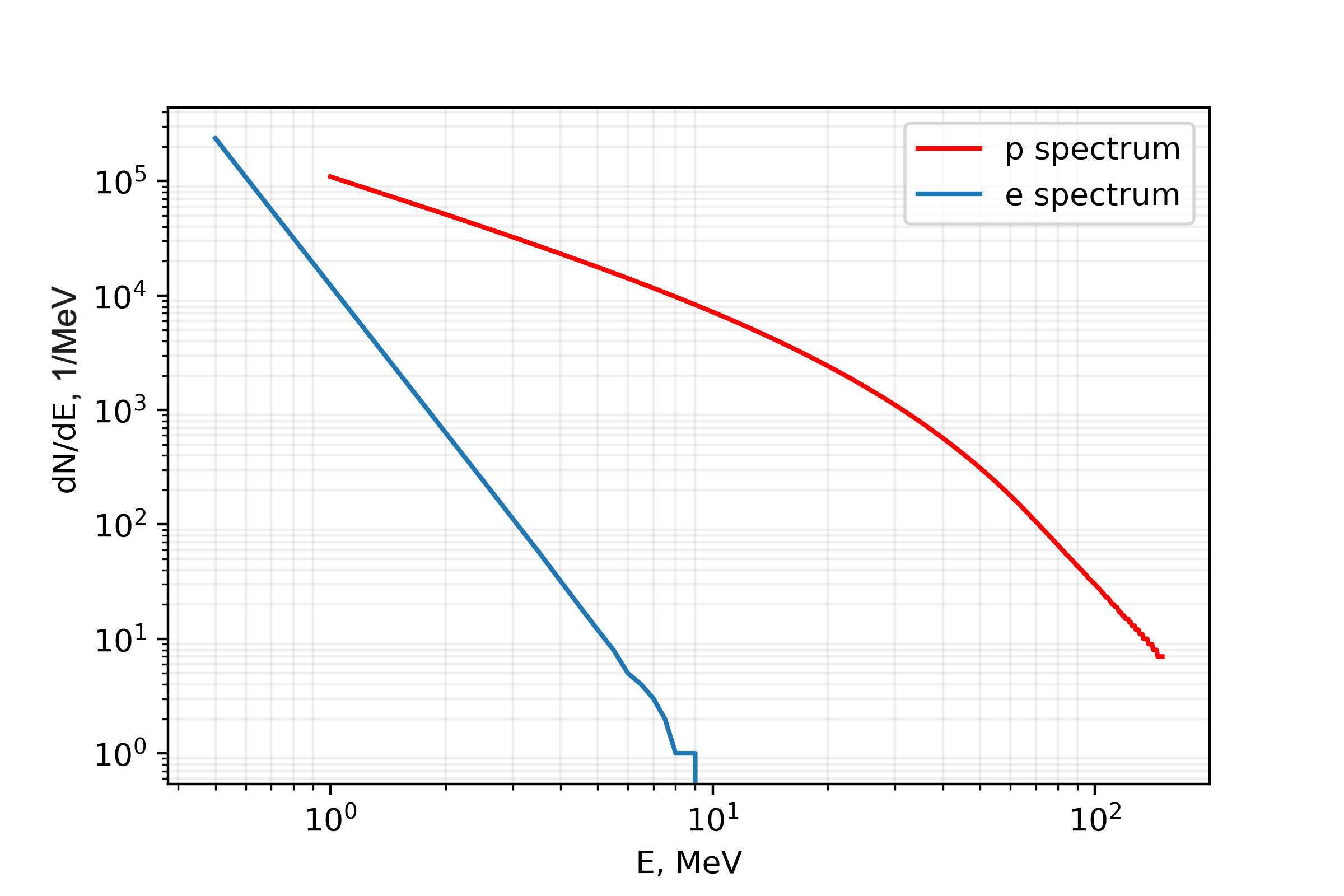}}
        \caption{The expected differential spectrum (total flux through the detector cross-section per section per 1 MeV) of protons and electrons calculated from \cite{real_energy_spectrum}.}
        \label{spectrum}
    \end{minipage}
    ~
    \begin{minipage}{0.5\linewidth}
        \center{\includegraphics[width=\textwidth]{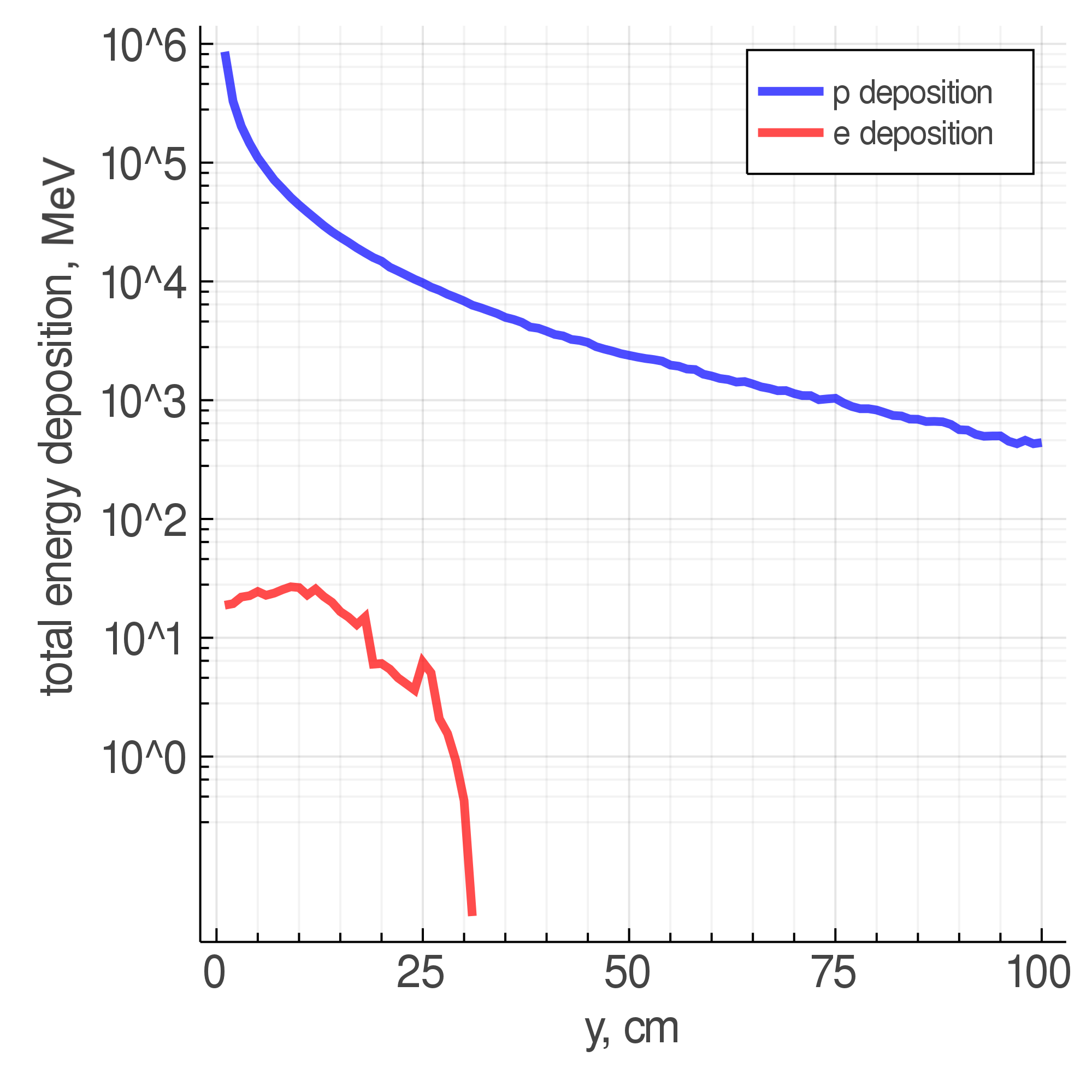}}
        \caption{Deposition of energy in the detector for spectrum shown at Fig.~\ref{spectrum}.}
        \label{deposition}
    \end{minipage}
\end{figure}

As mentioned above, the integral mode of the detector is required during periods of high SEP fluxes. As an example of data processing, we can take one of the SEP events from \cite{real_energy_spectrum}. The observed spectrum of protons and electrons is presented in Fig.~\ref{spectrum}.

The passage of particles through the detector is simulated using Geant4. The particle energy ranges were from 1 to 150 MeV for protons and from 0.5 to 15 MeV for electrons. We used both narrow and wide angular distribution for incident particles, but the results for both are quite similar since detector construction effectively cuts larger incident angles.

Turchin algorithm was implemented in Julia and Python programming languages. Python package was previously tested in the analysis of differential cross-section computation in the Troitsk nu-mass experiment \cite{nu_mass_data_analysis}. The new package -  TurchinReg.jl (\cite{turchinreg}) allows to use different bases and regularisation parameters. It also features advanced non-negativity boundary conditions, which limits the "loosening" of the spectrum (in this work it is not required).

From the form of the spectra (Fig. \ref{spectrum}) it is clear that in the integral mode the energy deposition of electrons will be much smaller than the energy emission of protons. It happens because the number of electrons is by orders of magnitude less than the number of protons for most energies and the energy deposition of electrons is about 10 times less for electrons with the same penetration length. The energy deposition for protons and electrons is shown in Fig.~\ref{deposition}. Based on this consideration, we used the combined realistic data including both protons and electrons, and reconstructed only the proton component.

\begin{figure}
        \center{\includegraphics[width=0.6\textwidth]{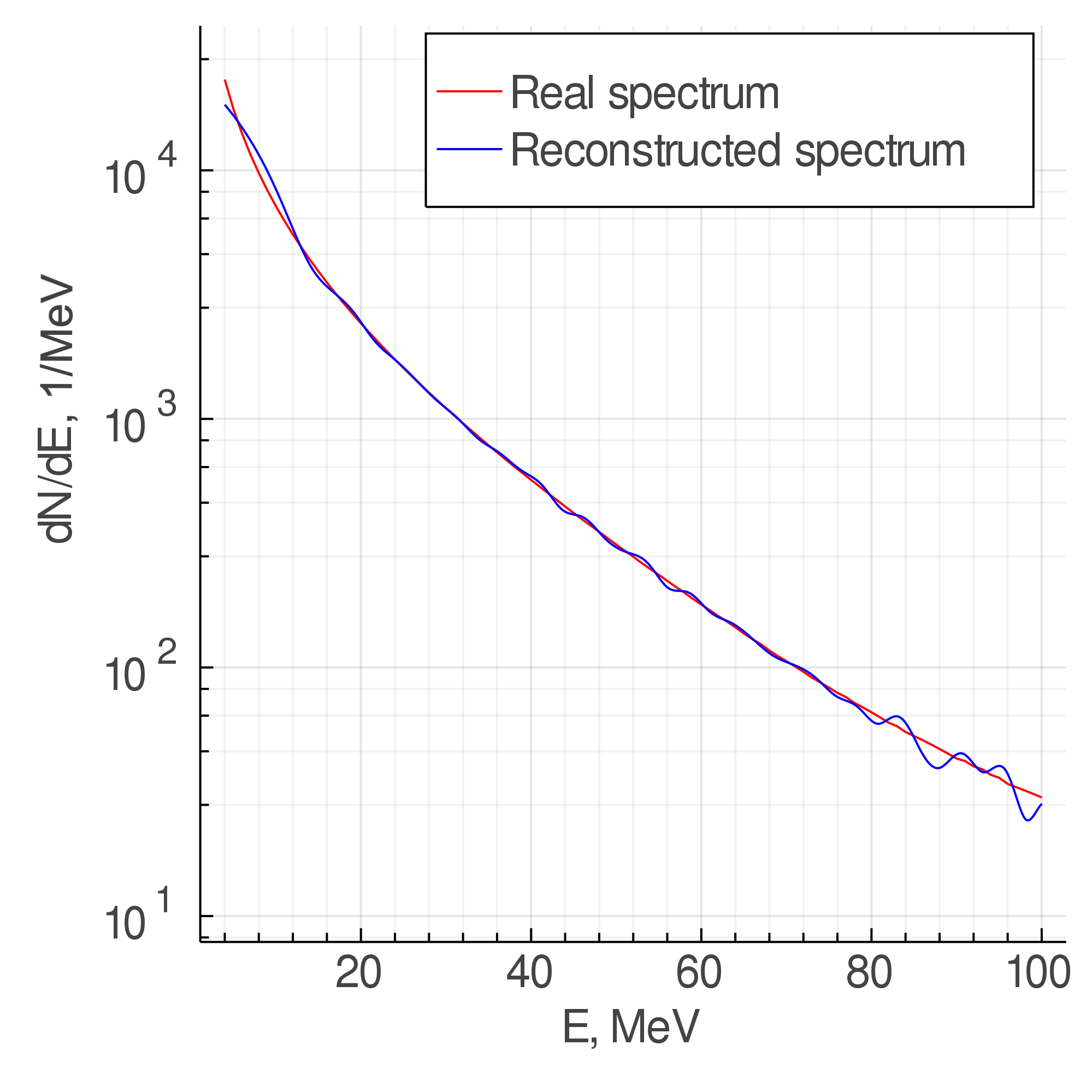}}
        \caption{Reconstruction of proton spectrum from the realistic mixed data shown at Fig~\ref{deposition}.}
        \label{solution_cont}
\end{figure}

The result of data processing is presented in figure \ref{solution_cont}. The relative accuracy of spectrum reconstruction using the Turchin algorithm is better than 7\% in the whole energy region and even better in its central part. Thus, the proton spectrum can be restored neglecting the contribution from electrons. We can conclude that the proton spectrum could be effectively reconstructed in the whole energy range without a loss of precision compared to the differential mode. On the other hand, it is not possible to get an electron spectrum from the integral mode unless the number of electrons is increased by 2-3 orders of magnitude, or the protons are somehow suppressed. 

An important feature of the integral mode is that it produces a very small amount of data - one number for each segment for the whole measurement time, which could last from seconds to minutes. It allows to send the whole amount of data to the earth and process it offline instead of burdening on-board computer.

With more complicated electronics it is possible to implement a hybrid mode, where forward segments (capturing the major part of low-energy particles) are working in integral mode and back segments are working in differential mode. More detailed simulations and reconstruction studies are required for specific satellite designs.

\section{Conclusion}

The detector prototype presented in this article has several key features:
\begin{itemize}
    \item Its construction is simple and it has a small mass. What is more important is that it does not require heavy shielding and collimators and effectively uses almost all its mass.
    \item The material of the detector is solid plastic, which makes it less susceptible to radiation and thermal damage, than semi-conductor detectors and less fragile than gas detectors.
    \item The segmentation allows us to significantly improve energy resolution in differential (individual event) measurement mode. The detection energy threshold could be adjusted by turning off frontal detector segments and thus allowing to use of differential mode measurements even during solar events when the count rate is high enough.
    \item Another benefit of segmentation is the ability to use integral (spectral measurement) mode in case of very high count rates ($\sim 10^{6}$ Hz and even higher) while limiting the amount of computation time and data transmission rates. 
    \item Integral mode analysis is supplemented by the unique implementation of Turchin's statistical regularization algorithm for spectrum reconstruction.
\end{itemize}

Laboratory tests and simulation results show its suitability for spectroscopy of selected particles. Despite relatively poor energy resolution of detector segments themselves, the profile reconstruction from all segments gives better than 5\% relative precision for protons and electrons in differential mode. Turchin's regularisation allows proton spectrum restoration with similar accuracy. Electrons could not be currently reconstructed in integral mode with given solar spectra, but it should be possible with the differential or hybrid mode.

In the future, it is planned to develop new electronics suitable for detector operation in space. Moreover, detector design will be adjusted to spacecraft requirements, e.g. minimization of mass-size characteristics, side shielding, etc. Also it is planned to develop thinner frontal segmentation for detector to allow precise measurement of electron spectrum around 1 MeV.

This work is supported by the Russian Science Foundation under grant No. 17-72-20134.

\section*{Appendix: Muon signal distribution}

\label{sec: Appendix}

In most cases, people use a symmetrical distribution shape to describe the amplitude spectrum of a scintillator detector. We found that it has a pronounced asymmetry that could not be explained by detector physics or electronics. We found out that the asymmetry could be attributed to the detector geometry.

\subsection*{Assumptions}
\begin{enumerate}
    \item The distribution of muons by the angle of incidence on the scintillator disk is $f(\theta) = cos^2\theta$ with $\theta \in \left[ 0, \frac{\pi}{2} \right]$
    \item Muons always fly through the top and the bottom of the disk or through both sides of the disk.
    \item Error for the fixed signal has normal distribution, but its standard deviation depends on the value of the signal.
    \item Disk light collection heterogeneity is ignored.
\end{enumerate}

\begin{figure}[ht] 
    \centering
    \includegraphics[width=0.7\linewidth]{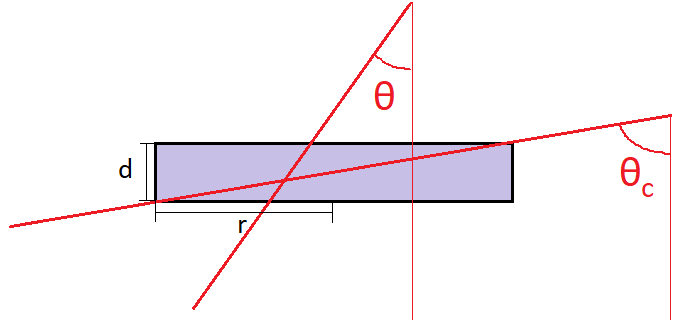}
    \caption{Muon propagation through detector channel scheme.}
    \label{1_1}
\end{figure}

\subsection*{Muon energy deposition}

One channel of the detector is a scintillator disk with a cylinder shape. The disk is made out of polystyrene. Its radius is $r =$ 3 cm and it's width is $d =$ 0.4 cm.

When a muon propagates through the disk perpendicular to the top surface disk it loses $\varepsilon_0 =$ 1.02 MeV within the scintillator. Muons are MIPs (Minimum Ionizing Particles), consequently, if a muon propagates through a layer of scintillator with length $= l$, then it loses $\varepsilon_0 \cdot \frac{l}{d}$ of its energy.

Muon propagates through the disk diagonal when its angle of incidence is equal to $\theta_{c} = arcctg(\frac{d}{2r}) \approx 1.44$ radian (Fig.~\ref{1_1}). This angle further will be called the critical angle. If muon's angle of incidence is less than $\theta_c$ then under made assumptions it will cross the disk through the top and the bottom surfaces. So its energy deposit can be found in the following way:

\begin{equation}
    \varepsilon(\theta) = \frac{\varepsilon_0}{\cos\theta}
\end{equation}

If muon's angle of incidence is more than $\theta_{c}$ then the muon propagates through side surface of the disk. Consequently, its energy deposit can be estimated in the following way:

\begin{equation}
    \varepsilon(\theta) = \varepsilon_0 \frac{2 r}{d \sin\theta}
\end{equation}

Minimum muon energy deposit is equal to $\varepsilon(0) = \varepsilon_0 = 1.02$ MeV. Maximum muon energy deposit is equal to $\varepsilon_c = \varepsilon(\theta_c) \approx 7.72$ MeV. And the general formula looks as follows:

\begin{displaymath}
\varepsilon(\theta) = \left\{ 
\begin{array}{ll}
    \frac{\varepsilon_0}{\cos\theta} & \textrm{$\theta \leq \theta_c$}\\
    \varepsilon_0 \frac{2 r}{d \sin\theta} & \textrm{$\theta \geq \theta_c$}
    \end{array} \right.
\end{displaymath}

\subsection*{Muon energy deposit distribution}

Let muon angle distribution be $f(\theta) = \cos^2\theta$. Let muon energy deposit in the channel distribution be $f(\varepsilon)$. Considering $\varepsilon$ being function of $\theta$ the following equation takes place:

\begin{equation}
    f(\theta) d\theta = f(\varepsilon) d\varepsilon = f(\varepsilon) \frac{d\varepsilon}{d\theta} d\theta
\end{equation}

$\varepsilon(\theta)$ consists of two parts. If $\theta \leq \theta_c$, then

\begin{equation}
    f(\varepsilon) \frac{d\varepsilon}{d\theta} d\theta = f(\varepsilon) \varepsilon_0 \frac{sin\theta}{cos^2\theta} d\theta = f(\theta) d\theta
\end{equation}

Consequently,

\begin{equation}
    f(\varepsilon) = f_0 \cdot cos^2\theta \cdot \frac{cos^2\theta}{\varepsilon_0 sin\theta} = f_0 \frac{\varepsilon_0^3}{\varepsilon^4 \sqrt{1 - (\frac{\varepsilon_0}{\varepsilon})^2}}
\end{equation}

Here $f_0$ is a normalization coefficient. Similarly for $\theta > \theta_c$ the distribution is as follows:

\begin{equation}
    f(\varepsilon) = f_0 cos^2(\theta) \frac{sin^2\theta}{\varepsilon_0 cos\theta} = f_0 \varepsilon_0 \left(\frac{2r}{\varepsilon d}\right)^2 \sqrt{1 - \left(\frac{2r\varepsilon_0}{\varepsilon d}\right)^2}
\end{equation}

Denoting $\varepsilon(\theta_c)$ as $\varepsilon_c$, the normalization coefficient is sought via following formula:

\begin{equation}
    f_0 = \frac{1}{\int_{\varepsilon_0}^{\varepsilon_{c}} \frac{\varepsilon_0^3}{\varepsilon^4 \sqrt{1 - (\frac{\varepsilon_0}{\varepsilon})^2}} d\varepsilon + \int_{\varepsilon(\frac{\pi}{2})}^{\varepsilon_{c}} \varepsilon_0 \left(\frac{2r}{\varepsilon d}\right)^2 \sqrt{1 - \left(\frac{2r\varepsilon_0}{\varepsilon d}\right)^2} d\varepsilon}
\end{equation}

\begin{figure}
    \begin{minipage}{0.5\linewidth}
        \centering
        \includegraphics[width=\linewidth]{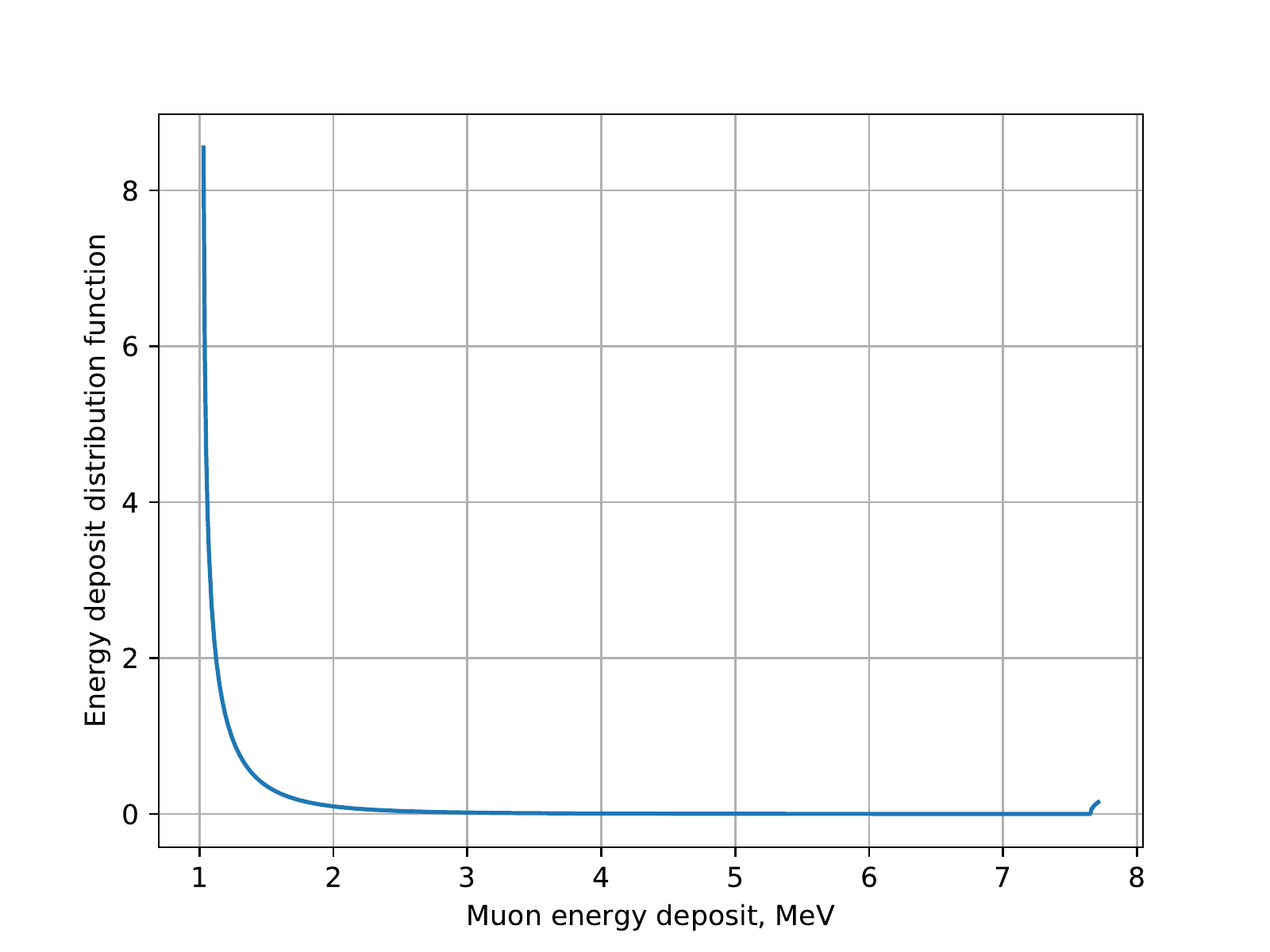}
        \caption{Muon energy deposit within a detector channel distribution.}
        \label{1_2}
    \end{minipage}
    ~
    \begin{minipage}{0.5\linewidth}
        \centering
        \includegraphics[width=\linewidth]{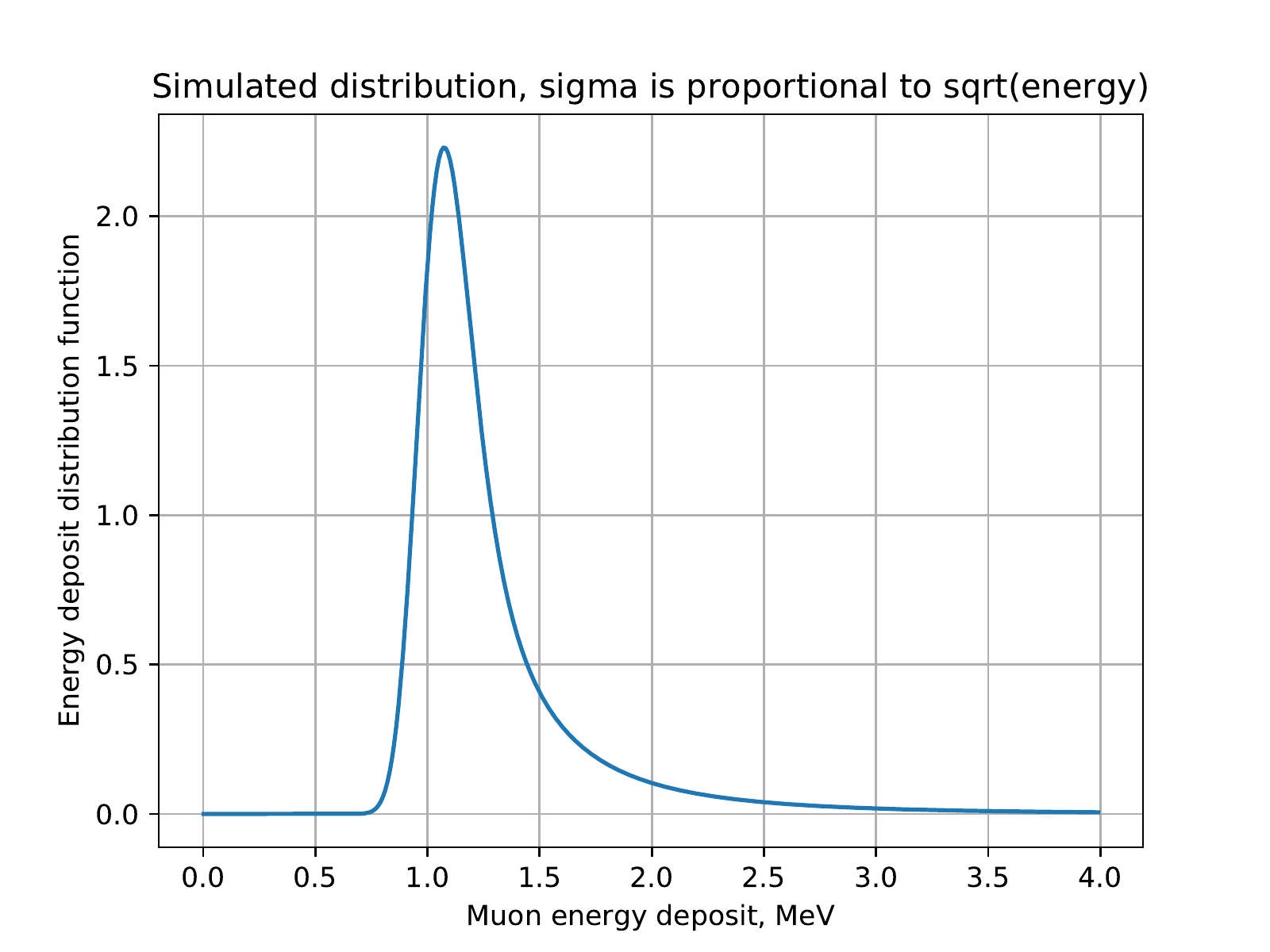}
        \caption{Example of theoretical muon signal distribution with good detector resolution.}
        \label{1_3}
    \end{minipage}
\end{figure}

$\varepsilon(\frac{\pi}{2}) \approx 7.65$ MeV $< \varepsilon_c$. Consequently, for $\varepsilon \in (7.65 MeV, 7.72 MeV)$ the total distribution is a sum of distributions for both considered cases. Therefore, muon energy deposit distribution is as follows (Fig.~\ref{1_2}):

\begin{displaymath}
f(\varepsilon) = \left\{ 
\begin{array}{ll}
    f_0 \cdot cos^2\theta \cdot \frac{cos^2\theta}{\varepsilon_0 sin\theta} = f_0 \frac{\varepsilon_0^3}{\varepsilon^4 \sqrt{1 - (\frac{\varepsilon_0}{\varepsilon})^2}} & \textrm{$\varepsilon <$ 7.65 MeV}\\
    f_0 cos^2\theta \frac{sin^2\theta}{\varepsilon_0 cos\theta} = f_0 \varepsilon_0 \left(\frac{2r}{\varepsilon d}\right)^2 \sqrt{1 - \left(\frac{2r\varepsilon_0}{\varepsilon d}\right)^2} & \textrm{$\varepsilon \in (7.65 MeV, 7.72 MeV)$}
    \end{array} \right.
\end{displaymath}

\begin{multline*}
    f_0^{-1} = \frac{1}{4}sin\left(2arccos(\frac{\varepsilon_0}{\varepsilon})\right)\big|_{\varepsilon_0}^{\varepsilon_{c}} + \frac{arccos(\frac{\varepsilon_0}{\varepsilon})}{2}\big|_{\varepsilon_0}^{\varepsilon_{c}} + \\ \frac{r}{d}arccos(\frac{\varepsilon_0}{\varepsilon})\big|_{\varepsilon(\frac{\pi}{2})}^{\varepsilon_{c}} - \frac{r}{2d}sin\left(2arccos(\frac{\varepsilon_0}{\varepsilon})\right)\big|_{\varepsilon(\frac{\pi}{2})}^{\varepsilon_{c}}
\end{multline*}

\subsection*{Muon signal distribution}

If a detector channel is irradiated with a monochromatic particle beam then every particle from this beam will deposit the same energy within the channel. But the signal will have a normal distribution due to the detector's signal statistical spread. Also, it is a well-known fact that scintillator detectors' errors are root-dependent on the energy deposit. Consequently, muon signal distribution is sought from the following convolution (Fig.~\ref{1_3}):

\begin{equation}
    F(\varepsilon) = \int_0^{\varepsilon_c} f(E) N(\varepsilon - E, \sigma \sqrt{\varepsilon}) dE
\end{equation}

Here $\sigma$ is a coefficient in the error root-dependence. Its physical meaning - the amplitude error for 1 MeV energy deposit.

%\addbibresource{bibliography.bib}
%\printbibliography
\bibliographystyle{unsrt}
\bibliography{bibliography}

\end{document}